\documentclass[conference]{IEEEtran}
\usepackage{graphicx}
\usepackage{amsmath,amssymb,dsfont,bbm,epstopdf,pgfplots,algorithm, algpseudocode} 
\usepackage{color}
\usepackage{cite}
\usepackage{graphics}
\usepackage{tikz}
\usepackage{pgfplots}
\usepackage{caption}
\usepackage{subcaption}
\usetikzlibrary{shapes, arrows, decorations.markings, arrows.meta}
\usetikzlibrary{decorations.markings}
\usetikzlibrary{patterns} 
\allowdisplaybreaks
\graphicspath{{.}{./Figures/}} 
\allowdisplaybreaks[4] 

\newtheorem{theorem}{Theorem}

\newcommand{\D}{\text{D}}
\newcommand{\Dt}{\D_t}
\newcommand{\Dr}{\D_r}
\renewcommand{\Pr}{{\mathbb{P}}}

\newcommand{\sRkf}{{\{1,\ldots,\lfloor 2^{n R_k^{(F)}} \rfloor \}}}
\newcommand{\sRks}{{\{1,\ldots,\lfloor 2^{n R_k^{(S)}} \rfloor \}}}

\newcommand{\mutx}{\mu_{\text {Tx}}}
\newcommand{\murx}{\mu_{\text {Rx}}}

\renewcommand{\S}{{\sf{S}}}

\newcommand{\Ntk}{\mathcal{N}_{\textnormal{Tx}}(k)}
\newcommand{\Nrk}{\mathcal{N}_{\textnormal{Rx}}(\tilde{k})}
\newcommand{\Iktilde}{\mathcal{I}(\tilde{k})}
\renewcommand{\i}{{\iota}}

\newcommand{\K}{[1:3K]}
\newcommand{\Ktilde}{[1:{K}]}

\newcommand{\vect}[1]{\boldsymbol{#1}}

\newcommand{\Tx}{\textnormal{Tx}}
\newcommand{\Rx}{\textnormal{Rx}}

\newcommand{\MkF}{M_k^{(F)}}
\newcommand{\MkS}{M_k^{(S)}}

\definecolor{ForestGreen}{rgb}{0.0, 0.5, 0.0}

\newcommand{\fillaroundd}[3][fill=green!30]{
\node[filledhex,#1] at (#3,#3+4) {};
}

\begin{document}
\title{Multiplexing Gain Region of Sectorized Cellular Networks with Mixed Delay Constraints}
\author{\IEEEauthorblockN{Homa Nikbakht and Mich\`ele Wigger}
	\IEEEauthorblockA{ LTCI, T$\acute{\mbox{e}}$l$\acute{\mbox{e}}$com ParisTech\\ 
		\{homa.nikbakht, michele.wigger\}@telecom-paristech.fr}
	\and
	\IEEEauthorblockN{Shlomo Shamai (Shitz)}
	\IEEEauthorblockA{Technion\\
		sshlomo@ee.technion.ac.il}}

\maketitle

\IEEEpeerreviewmaketitle

\begin{abstract}
The sectorized hexagonal model with mixed delay constraints  is considered when both neighbouring mobile users and base stations can  cooperate over rate-limited  links. Each message is a combination of independent ``fast'' and ``slow'' bits, where the former are subject to a stringent delay constraint  and cannot profit from cooperation. Inner and outer  bounds on the multiplexing gain region are derived. The obtained results show that for small cooperation prelogs or moderate ``fast'' multiplexing gains, the overall performance (sum multiplexing gain) is hardly decreased by the stringent delay constraints on the ``fast" bits. For large cooperation prelogs and large ``fast'' multiplexing gains, increasing the ``fast'' multiplexing gain by $\Delta$ comes at the expense of decreasing the sum multiplexing gain by $3 \Delta$.
%
%

\end{abstract}

\section{Introduction}
Heterogenous traffics bring various challenges for modern communication systems.  In this paper, we consider traffics with heterogeneous latency requirements. Specifically, we consider a scenario where  communication for  delay-tolerant applications can exploit cooperation between terminals, but communication for delay-sensitive applications cannot.  
Previous works on  wireless systems with mixed delay constraints include \cite{ shlomo2012ISIT, Zhang2005, Zhang2008, Kassab2018, long, HomaITW2018, Kassab2019, Ghanem2019, Wiggeretal}. In particular,  \cite{shlomo2012ISIT} proposes a broadcasting approach over a single-antenna fading channel to   communicate a stream of ``fast" messages, which have to be  sent over a single coherence block, and a stream of ``slow" messages, which can be sent over   multiple blocks.  On the other hand, \cite{Zhang2005} time-shares ``fast'' and ``slow'' messages over different frames, where the stringent delay constraint prevents employing sophisticated power and rate allocation strategies on the transmission of ``fast'' messages. A scheduling algorithm that prioritizes ``fast'' messages over ``slow'' messages is proposed in \cite{Zhang2008}. Mixed delay constraints in the context of cloud radio access networks (C-RAN) where ``fast'' messages need to be decoded directly at the base stations are studied in \cite{Kassab2018}.

 In  \cite{long,HomaITW2018} the multiplexing gain and capacity regions of Wyner's soft-handoff network \cite{Wyner-94,Hanly-Whiting-93,Simeone-2011-Tut} under mixed delay constraints are studied. The works  \cite{long,HomaITW2018}   show that when only the transmitters or only the receivers can cooperate, and when the rate of ``fast'' messages is only moderate, then the sum-rate is not decreased compared to a scenario where only  ``slow'' messages are transmitted. In contrast, when  ``fast'' messages are of high rates, $1$ bit of  ``fast'' messages  comes at the expense  of $2$ bits of ``slow'' messages. 
The situation is different, when both the transmitters \emph{and} the receivers can cooperate and cooperation rates are sufficiently large. Then, through sophisticated coding schemes the maximum  \emph{sum} multiplexing gain is achievable at any delay-sensitive rate.  In such a scenario, the stringent delay constraints thus do not harm the system's overall performance (sum multiplexing gain).

A similar conclusion was obtained in \cite{HomaITW2018} for the  sectorized hexagonal model \cite{Katz2013, Samet, Khina2016} under mixed delay constraints when   both the mobile users and the base stations can cooperate. More precisely, \cite{HomaITW2018} proposed a lower bound on the multiplexing gain region when the capacity of the cooperation links is unconstrained. The maximum sum multiplexing gain of this lower bound can even be achieved.

 In this paper, we extend the result in \cite{HomaITW2018} to the practically more relevant setup where the cooperation links between mobile users and between basestations are rate-constrained. We also prove a converse bound on the   multiplexing gain region of the sectorized hexagonal model with mixed delay constraints and rate-constrained cooperation links. For small cooperation rates the sum multiplexing gain does not decrease even for large delay-sensitive rates. The stringent delay constraint thus again does not degrade the system's overall performance.

\section{Problem Setup}
 \begin{figure}[t]
  \vspace*{-2ex}
  	\centering
  	\small
  	\begin{tikzpicture}[scale=0.8, >=stealth]
  	\centering
  	
  	  \draw[fill=gray!50] (2+3,1.7321+2*1.7321) -- (2+3,1.7321+2*1.7321+0.8660) --(3*1+3*cos{60},2*sin{60}*3+sin{60})--(3*1+1,2*sin{60}*3)--+(-60:0.5)--(2+3,1.7321+2*1.7321) ;
  	  \draw [fill = pink!60] (0.5+3*1,0.8660+3*1.7321)-- +(-30:0.8660)--(3*1+1,2*sin{60}*3)--(3,2*sin{60}*3) --+(120:0.5)--(0.5+3*1,0.8660+3*1.7321);
  	  \draw [fill =pink!60] (0.5+3*1,0.8660+3*1.7321) -- (0.5+3*1,0.8660+3*1.7321+ 0.8660 )-- (0.5+3*1 + 0.5,0.8660+3*1.7321+ 0.8660 )--(3*1+3*cos{60},2*sin{60}*3+sin{60})--+(-120:0.5)-- (0.5+3*1,0.8660+3*1.7321);
  	  
  	   \draw [fill =pink!60] (0.5+3*1,0.8660+3*1.7321 -1.7321 ) -- (0.5+3*1,0.8660+3*1.7321+ 0.8660-1.7321 )-- (0.5+3*1 + 0.5,0.8660+3*1.7321+ 0.8660-1.7321 )--(3*1+3*cos{60},2*sin{60}*3+sin{60}-1.7321)--+(-120:0.5)-- (0.5+3*1,0.8660+3*1.7321-1.7321);
  	   
  	  \draw [fill = pink!60] (0.5+3*1+1.5,0.8660+3*1.7321+0.8660)-- +(-30:0.8660)--(3*1+1+1.5,2*sin{60}*3+0.8660)--(3+1.5,2*sin{60}*3+0.8660) --+(120:0.5)--(0.5+3*1+1.5,0.8660+3*1.7321+0.8660);
  	   


  	 \foreach \i in {2} 
  \foreach \j in {2,3,4} 
   \foreach \jj in {2,3} {
  \foreach \a in {0,120,-120} \draw[black] (3*\i,2*sin{60}*\j) -- +(\a:1);
  \foreach \a in {90,-30,-150} \draw[black,dashed] (0.5+3*\i,0.8660+\jj*1.7321) -- +(\a:0.8660);
  \draw [fill=black] (0.5 + 3*\i, 1.7321/4+1.7321*\jj) circle (0.07);
  \draw [fill=black] (0.5+0.5*0.75+ 3*\i, 1.7321/2+1.7321/8+1.7321*\jj) circle (0.07);
  \draw [fill=black] (0.5-0.5*0.75+ 3*\i, 1.7321/2+1.7321/8+1.7321*\jj) circle (0.07);
   }
  \foreach \i in {1} 
  \foreach \j in {3} {
  \foreach \a in {0,120,-120} \draw[black] (3*\i,2*sin{60}*\j) -- +(\a:1);
  \foreach \a in {90,-30,-150} \draw[black,dashed] (0.5+3*\i,0.8660+\j*1.7321) -- +(\a:0.8660);
    \draw [fill=black] (0.5 + 3*\i, 1.7321/4+1.7321*\j) circle (0.07);
    \draw [fill=black] (2+ 3*\i, 3*1.7321/4+1.7321*\j) circle (0.07);
      \draw [fill=black] (0.5+0.5*0.75+ 3*\i, 1.7321/2+1.7321/8+1.7321*\j) circle (0.07);
  \draw [fill=black] (0.5-0.5*0.75+ 3*\i, 1.7321/2+1.7321/8+1.7321*\j) circle (0.07);}
  \foreach \i in {1} 
  \foreach \j in {2} {
  \foreach \a in {0,120} \draw(3*\i,2*sin{60}*\j) -- +(\a:1);
   \foreach \a in {90,-30,-150} \draw[black,dashed] (0.5+3*\i,0.8660+\j*1.7321) -- +(\a:0.8660);
     \draw [fill=black] (0.5 + 3*\i, 1.7321/4+1.7321*\j) circle (0.07);
     \draw [fill=black] (2+ 3*\i, 3*1.7321/4+1.7321*\j) circle (0.07);
       \draw [fill=black] (0.5+0.5*0.75+ 3*\i, 1.7321/2+1.7321/8+1.7321*\j) circle (0.07);
  \draw [fill=black] (0.5-0.5*0.75+ 3*\i, 1.7321/2+1.7321/8+1.7321*\j) circle (0.07);}
    \foreach \i in {1} 
  \foreach \j in {4} {
  \foreach \a in {0,-120} \draw(3*\i,2*sin{60}*\j) -- +(\a:1);
}
  
    \foreach \i in {1} 
  \foreach \j in {1,2,3} {
  \draw [fill=black] (2+ 3*\i, 3*1.7321/4+1.7321*\j) circle (0.07);
   \draw [fill=black] (2-0.5*0.75+3*\i, 1.7321/2+1.7321/2+1.7321/8+1.7321*\j) circle (0.07);
   \draw [fill=black] (2+0.5*0.75+3*\i, 1.7321/2+1.7321/2+1.7321/8+1.7321*\j) circle (0.07);}
   \foreach \i in {1} 
  \foreach \j in {2,3} 
   \foreach \jj in {1,2,3} {
 \foreach \a in {0,120,-120} \draw[black](3*\i+3*cos{60},2*sin{60}*\j+sin{60}) -- +(\a:1);
  \foreach \a in {90,-30,-150} \draw[black,dashed] (2+3*\i,1.7321+\jj*1.7321) -- +(\a:0.8660);}

  \foreach \i in {1} 
  \foreach \j in {4} {
 \foreach \a in {0,-120} \draw(3*\i+3*cos{60},2*sin{60}*\j+sin{60}) -- +(\a:1);}
  \foreach \i in {1} 
  \foreach \j in {1} {
 \foreach \a in {0,120} \draw(3*\i+3*cos{60},2*sin{60}*\j+sin{60}) -- +(\a:1);}
  \foreach \i in {2} 
  \foreach \j in {2,3} {
 \foreach \a in {120,-120} \draw (3*\i+3*cos{60},2*sin{60}*\j+sin{60}) -- +(\a:1);}
 
    \foreach \cc in {2,3,4}
     \foreach \c in {1,2}{
 \draw [blue] (0.5+3*\c , 1.7321/4+1.7321*\cc) --(0.5+0.5*0.75+ 3*\c, 1.7321/2+1.7321/8+1.7321*\cc- 1.7321);
 \draw [blue] (0.5+3*\c , 1.7321/4+1.7321*\cc) --(0.5-0.5*0.75+ 3*\c, 1.7321/2+1.7321/8+1.7321*\cc- 1.7321);}
 \foreach \cc in {2,3}
     \foreach \c in {2}{
 \draw [blue] (0.5+3*\c , 1.7321/4+1.7321*\cc) --(2-0.5*0.75+3*\c, 1.7321/2+1.7321/2+1.7321/8+1.7321*\cc- 1.7321);}
  \foreach \cc in {2,3,4}
     \foreach \c in {1}{
      \draw [blue] (0.5+3*\c , 1.7321/4+1.7321*\cc) --(2-0.5*0.75+3*\c, 1.7321/2+1.7321/2+1.7321/8+1.7321*\cc- 1.7321);
 \draw [blue] (2+0.5*0.75+3*\c, 1.7321/2+1.7321/2+1.7321/8+1.7321*\cc- 1.7321) -- (0.5+3*\c +3, 1.7321/4+1.7321*\cc);}
  \foreach \cc in {2,3}
     \foreach \c in {1}{
  \draw [blue] (0.5+3*\c , 1.7321/4+1.7321*\cc) --(2+0.5*0.75+3*\c-3, 1.7321/2+1.7321/2+1.7321/8+1.7321*\cc- 1.7321);}
  
  \foreach \cc in {1,2,3}
     \foreach \c in {1}{
  \draw [blue] (2+3*\c , 3*1.7321/4+1.7321*\cc) --(0.5+0.5*0.75+ 3*\c, 1.7321/2+1.7321/8+1.7321*\cc);}
  
   \foreach \cc in {1,2,3}
     \foreach \c in {1}{
 \draw [blue] (2+3*\c , 3*1.7321/4+1.7321*\cc) --(0.5-0.5*0.75+ 3*\c+3, 1.7321/2+1.7321/8+1.7321*\cc);}

   \foreach \cc in {1,2,3,4}
     \foreach \c in {1}{
 \draw [blue] (2+3*\c , 3*1.7321/4+1.7321*\cc) --(2-0.5*0.75+3*\c, 1.7321/2+1.7321/2+1.7321/8+1.7321*\cc- 1.7321);
  \draw [blue] (2+3*\c , 3*1.7321/4+1.7321*\cc) --(2+0.5*0.75+3*\c, 1.7321/2+1.7321/2+1.7321/8+1.7321*\cc- 1.7321);}

 \foreach \cc in {2,3,4}
     \foreach \c in {1}{  
   \draw [blue] (0.5+0.5*0.75+3*\c, 1.7321/2+1.7321/8+1.7321*\cc) -- (2-0.5*0.75+ 3*\c, 1.7321/2+1.7321/2+1.7321/8+1.7321*\cc- 1.7321);}
    \foreach \cc in {2,3}
    \foreach \c in {2}{  
   \draw [blue] (0.5+0.5*0.75+3*\c, 1.7321/2+1.7321/8+1.7321*\cc) -- (2-0.5*0.75+ 3*\c, 1.7321/2+1.7321/2+1.7321/8+1.7321*\cc- 1.7321);}
    \foreach \cc in {1,2,3}
     \foreach \c in {1}{ 
    \draw [blue] (0.5+0.5*0.75+3*\c, 1.7321/2+1.7321/8+1.7321*\cc) -- (2-0.5*0.75+ 3*\c, 1.7321/2+1.7321/2+1.7321/8+1.7321*\cc);}
    
    \foreach \cc in {1,2,3}
     \foreach \c in {1}{ 
     \draw [blue] (2+0.5*0.75+3*\c, 1.7321/2+1.7321/2+1.7321/8+1.7321*\cc) -- (0.5-0.5*0.75+ 3+3*\c, 1.7321/2+1.7321/8+1.7321*\cc);}
     \foreach \cc in {1,2,3}
     \foreach \c in {1}{
     \draw [blue] (2+0.5*0.75+3*\c, 1.7321/2+1.7321/2+1.7321/8+1.7321*\cc) -- (0.5-0.5*0.75+ 3+3*\c, 1.7321/2+1.7321/8+1.7321+1.7321*\cc);}
     
       \foreach \cc in {1,2}
     \foreach \c in {0}{
     \draw [blue] (2+0.5*0.75+3*\c, 1.7321/2+1.7321/2+1.7321/8+1.7321*\cc) -- (0.5-0.5*0.75+ 3+3*\c, 1.7321/2+1.7321/8+1.7321+1.7321*\cc);}

  	\end{tikzpicture} 
  	\caption{Illustration of the sectorized hexagonal   network. Small circles indicate mobile users,  black solid lines depict the cell borders, dashed black lines the sector borders, and solid blue lines indicate that the communication in two given sectors interfere. }
  	\label{fig1-1}
  	\vspace*{-4ex}
  \end{figure}
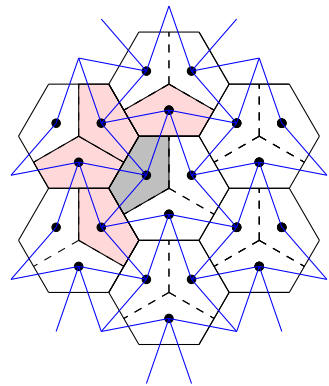~~%
Consider the uplink communication of a cellular network with ${K}$ cells, where each cell consists of three sectors. For simplicity, and because the focus is on the multiplexing gain, we assume a single mobile user in each sector. Each Tx (mobile user)  is associated with a distinct sector and is equipped with $M$ antennas.  Each Rx (base station) is associated with a distinct cell and is equipped with $3 M$ directional antennas, where  $M$ antennas are pointing to each of the three sectors of the cell.  This allows avoiding interference between communications from different sectors in the same cell. The interference pattern of our network  is depicted by the blue solid lines in Fig.~\ref{fig1-1},  where the three sectors of a cell are separated by dashed lines. Notice for example, that transmission in the gray shaded sector is interfered only by the transmissions in the four adjacent pink shaded sectors. We call the transmitters in the pink sectors the neighbourhood of the transmitter in the gray sector. More generally, for each Tx~$k \in \K$, the \emph{neighbourhood $\Ntk$} denotes the set of Txs whose communication interferes with Tx~$k$. 
The  time-t signal received at the $M$ receive antennas directing to Tx~$k$ can  then be written as:
\begin{equation}\label{Eqn:Channel}
\vect{Y}_{\tilde{k},t} = \mathsf{H}_{k,k}\vect{X}_{k,t} +\sum_{\ell \in \Ntk } \mathsf{H}_{k,\ell }\vect{X}_{\ell,t} +\vect{Z}_{k,t},
\end{equation}
where  the $M$-by-$M$   matrix $\mathsf{H}_{k,\ell}$ models the channel from mobile user $k$ to the receiving antennas in Sector $\ell$ and is randomly drawn according to a given continuous distribution.

 Each Tx~$k$ wishes to send a pair of independent  messages  $M_k^{(F)}$ and $M_k^{(S)}$ to a specific Rx~$\tilde{k}$ with $\tilde k \in [1: K]$. The ``fast"  message $\MkF$ is uniformly distributed over the set $\mathcal{M}_{k}^{(F)}:=\sRkf$ and needs to be decoded subject to a  stringent delay constraint, as we explain shortly.  The ``slow"  message $\MkS$ is uniformly distributed over $\mathcal{M}_{k}^{(S)}:=\sRks$ and is subject to a less stringent decoding delay constraint.  Here, $n$ denotes the blocklength of transmission and $R_k^{(F)}$ and $R_k^{(S)}$  the rates of transmissions of the ``fast" and ``slow" messages.   
\par Neighbouring Txs  cooperate during $\Dt>0$ rounds over dedicated noise-free, but rate-limited links and neighbouring Rxs  cooperate during $\Dr>0$ rounds. The cooperative communication  is subject to a total delay constraint 
\begin{equation}
\D_t +\D_r \leq \D.\label{eq:delay}
\end{equation}
where $\D>0$ is a given parameter of the system. As we will see, the stringent delay constraints on ``fast" messages impose limitations on the  cooperation. In particular, cooperation between Txs can only depend on ``slow" but not on ``fast" messages. 

We  describe the encoding at the Txs.  In each conferencing round $j\in\{1,\ldots, \Dt\}$,  Tx~$k\in \K$, produces  a conferencing message  $T_{k \to \ell}^{(j)}$ for each of its neighbours $\ell\in \Ntk$ by computing
	\begin{equation}\label{eq:txconf}
	T_{k\to \ell}^{(j)}  = \xi_{k\to \ell}^{(n)} \Big(M_{k}^{(S)}, \big\{T_{\ell' \to k}^{(1)}, \ldots, T_{\ell'\to k}^{(j-1)} \big\}_{\ell' \in \Ntk}\Big), 
	\end{equation}
	 for some function $\xi_{k \to \ell}^{(n)}$ on appropriate domains. Tx~$k$  sends the  messages $T_{k\to \ell}^{(1)}, \ldots, T_{k\to \ell}^{(\Dt)}$  over the conferencing link to Tx~$\ell$. The rate-limitation on the conferencing link imposes
	\begin{equation}\label{eq:conference_capatx}
	\sum_{j=1}^{\Dt}        H( T^{(j)}_{k\rightarrow  \ell})         \leq  \mu_{\Tx} \frac{n}{2} \log (P), \qquad k\in\K,\; \ell \in \Ntk,
	\end{equation}
	for a given $\mu_{\Tx}>0$. 

 Tx~$k$ finally computes its channel inputs as a function of  its ``fast" and ``slow" messages and of all the $\Dt \cdot |\Ntk|$ conferencing messages that it obtained from its neighbouring Txs:
\begin{equation}\label{xkn}
X_k^n  =  {f}_k^{(n)} \Big( M_{k}^{(F)}, M_{k}^{(S)}, \{T_{\ell' \to k}^{(1)}, \ldots, T_{\ell'\to k}^{(\Dt)} \}_{\ell' \in \Ntk} \Big).
\end{equation}


 The channel inputs have to satisfy the average block-power constraint
\begin{equation}\label{eq:power}
\frac{1}{n} \sum_{t=1}^n X_{k,t}^2
\leq P, \quad \text{a.s.},
\quad \forall\ k \in \K.
\end{equation}

We now describe the decoding. For each $\tilde{k}\in\Ktilde$, let $\Iktilde$ denote the set of messages intended for a given Rx~$\tilde{k}$, and let $\Nrk$ denote the set of Rxs that are neighbours of Rx~$\tilde{k}$. Decoding takes place in two phases. During the first \emph{fast-decoding phase}, each  Rx~$\tilde{k}$  decodes all its intended  ``fast"  messages $\big\{ M_\ell^{(F)} \colon \ell \in \Iktilde\big\}$ based on its own channel outputs. So, it produces:
\begin{equation}\label{mhatf}
\hat{{\mathbf{M}}} _{\tilde{k}}^{(F)} ={g_{\tilde{k}}^{(n)}}\big( Y_{\tilde{k}}^{n}\big)
\end{equation} 
where $\hat{{\mathbf{M}}} _{\tilde{k}}^{(F)} := (\hat{{{M}}} _{\ell}^{(F)}\colon \ell \in \Iktilde)$ and where $g_{\tilde{k}}^{(n)}$ denotes a decoding function on appropriate domains.

In the subsequent \emph{slow-decoding phase},  Rxs first communicate with their neighbours  during $\Dr>0$ rounds. 
In each conferencing round $j'\in [1:\Dr]$,  each Rx~$\tilde{k}$, for $\tilde{k}\in[1:{K}]$, produces  a conferencing message  $Q_{\tilde{k} \to \ell}^{(j')}$ for each of its neighbours $\ell\in \Nrk$: 
\begin{equation}\label{eq:rxconf}
Q^{(j')}_{\tilde{k}\rightarrow \ell}={\psi_{\tilde{k} \to \ell}^{(n)}}\Big( Y_k^n,\big\{ Q^{(1)}_{\ell' \rightarrow \tilde{k}},\ldots , Q^{(j'-1)}_{\ell' \rightarrow \tilde{k}}\}_{\ell' \in \Nrk}\big\}\Big),  
\end{equation}
for an   encoding function $\psi_{\tilde{k} \to \ell}^{(n)}$  on appropriate domains. Rx~$\tilde{k}$ then sends the messages $Q^{(1)}_{\tilde{k}\rightarrow \ell},\ldots,$ $Q^{(\Dr)}_{\tilde{k}\rightarrow \ell}$ over the conferencing link to Rx~$\ell$. The rate-limitation on the conferencing link imposes that for a  given $\mu_{\Rx}>0$.
\begin{equation}\label{eq:conference_capa}
\sum_{j'=1}^{\Dr}	H( Q^{(j')}_{\tilde{k}\rightarrow \ell })	 \leq   \mu_{\Rx} \frac{n}{2} \log (P), \qquad \tilde{k} \in [1:{K}], \; \ell \in \Nrk.
\end{equation}

After the last conferencing round, each Rx~$\tilde{k}$ decodes its desired ``slow" messages as
\begin{equation}\label{mhats}
\hat{\mathbf{M}}_{\tilde{k}}^{(S)}={b_{\tilde{k}}^{(n)}}\Big( Y_{\tilde{k}}^n, \Big\{ Q^{(1)}_{\ell'\rightarrow \tilde{k}}, \ldots, Q^{(\Dr)}_{\ell'\rightarrow \tilde{k}}\Big\}_{\ell' \in \Nrk}\Big) 
\end{equation}
where $\hat{\mathbf{M}}_{\tilde{k}}^{(S)}:=( \hat{M}_{\ell}^{(S)} \colon \ell \in \Iktilde)$ and where  $b_{\tilde{k}}^{(n)}$ denotes a decoding function on appropriate domains.

%

Given cooperation prelogs  $\mu_{\Rx},\mu_{\Tx}\geq 0$ and maximum  delay $\D$, a multiplexing-gain pair  $(\S^{(F)},\S^{(S)})$ is called \emph{achievable}, if for every positive integer $K$  there exists a sequence  $\{R_K^{(F)}(P),R_K^{(S)}(P) \}_{P>0}$  so that 
\begin{align} \label{sf}
\S^{(F)}&:= \varlimsup_{K\rightarrow \infty}\; \varlimsup_{P\rightarrow\infty} \;\frac{  R_K^{(F)}(P)}{\frac{1}{2}\log (1+P)},\\\label{ss}
\S^{(S)}&:= \varlimsup_{K\rightarrow \infty}\; \varlimsup_{P\rightarrow\infty} \;\frac{  R_K^{(S)}(P)}{\frac{1}{2}\log (1+P)},
\end{align}
and  so that for each rate pair $(R_K^{(F)}(P),R_K^{(S)}(P))$ it is possible to find a sequence (in the blocklength $n$) of encoding, cooperation, and decoding functions  satisfying constraints \eqref{eq:delay},   \eqref{eq:conference_capatx}, \eqref{eq:power}, and  \eqref{eq:conference_capa} and with vanishing probability of error: 
\begin{equation}
p(\textnormal{error}) := \Pr \bigg[\bigcup_{k \in \K} \Big(\big( \hat{M}_k^{(F)} \neq M_k^{(F)}\big)  \cup \big(\hat{M}_k^{(S)} \neq M_k^{(S)}\big)\Big)  \bigg] 
\end{equation}
 goes to $0$ as $n$ goes to infinity. The  closure of the set of all achievable   multiplexing-gain pairs $(\S^{(F)}, \S^{(S)})$ is called \emph{multiplexing-gain region} and is denoted by $\mathcal{S}^\star(\mu_{\Tx},\mu_{\Rx},\D)$.

\section{Main Results}
\begin{theorem}\label{lemma1} 
	Given $\mutx$, $\murx$ and $\D$, the convex hull of the set containing the following multiplexing gain pairs is achievable: 
	\begin{IEEEeqnarray}{l} \hspace{2cm}(\S^{(F)}=0,\; \S^{(S)}=0)\\
		\hspace{2cm}(\S^{(F)}=\frac{M}{2},\; \S^{(S)}= 0)\\
		\bigg\{\hspace{-0.1cm}\Big(\S^{(F)} =0 , \nonumber \\ \hspace{.2cm} \S^{(S)}= \min\Big\{ \frac{M(3t-1)}{3t}, \frac{M}{2} + \frac{(\mutx + \murx) (3t-2)}{2t(2t-1)} \Big\}\Big)\hspace{-0.1cm}\bigg\}_{t=1}^{ \frac{\D}{4}}\nonumber \\ \label{s1} \\
			\bigg\{\Big(\S^{(F)} =0 , \nonumber \\
			 \hspace{.5cm} \S^{(S)}= \min\Big\{ \frac{M(3t-1)}{3t}, \frac{M}{2} + \frac{\murx (3t-2)}{2t(2t-1)} \Big\}\Big)\bigg\}_{t= \frac{\D}{4}+1 }^{\frac{\D}{2}}\nonumber \\ \label{s2} \\
		\bigg\{\Big(\S^{(F)} =\min \Big \{ \frac{M}{3}, \frac{M}{2} - \frac{3(\mutx + \murx) t^2}{(4t^2-1)(2t+ 3)}\Big\}, \nonumber \\ \hspace{1cm} \S^{(S)}= \min \Big \{\frac{M(2t -1)}{3t}, \frac{6 (\mutx+\murx) t}{4t^2+ 8t+ 3}\Big \} \Big)\bigg \}_{t=1}^{ \frac{\D-2}{4}} \label{s3} \\
		\bigg\{\Big(\S^{(F)} =\min \Big \{ \frac{M}{3}, \frac{M}{2} - \frac{3\murx t^2}{(4t^2-1)(2t+ 3)}\Big\}, \nonumber \\ \hspace{1cm} \S^{(S)}=  \min \Big \{\frac{M(2t -1)}{3t}, \frac{6 \murx t}{4t^2+ 8t+ 3}\Big \} \Big)\bigg\}_{t= \frac{\D+2}{4} }^{\frac{\D-2}{2}}.\label{s4} 
\end{IEEEeqnarray}
 \end{theorem}
 \begin{IEEEproof}
See Section~\ref{sec:ach.th1}.
  \end{IEEEproof}
 
\begin{figure}[!t]
\centering
\begin{tikzpicture}[scale=0.9]
\begin{axis}[
    xlabel={\small {$\S^{(F)}$ }},
    ylabel={\small {$\S^{(S)}$ }},
     xlabel style={yshift=.5em},
     ylabel style={yshift=-1.25em},
    xmin=0, xmax=1.6,
    ymin=0, ymax=4,
   xtick={0,0.5,1,1.5},
    ytick={0,1,1.5,2,2.5, 3},
    yticklabel style = {font=\small,xshift=0.25ex},
    xticklabel style = {font=\small,yshift=0.25ex},
]
 \addplot[ color=black] coordinates { (0,2.9964)(1.5,1.4964)(1.5,0) };
\addplot[ color=blue] coordinates { (0,2.75)(1,1.75)(1.5,0) };
    \addplot[ color=black, densely dash dot, thick] coordinates { (0,1.7667)(1.5,0.2667)(1.5,0) };
\addplot[color=blue, densely dash dot dot, thick] coordinates {(0,1.5536)(1.4792,0.0727)(1.5,0) };

\legend{{\footnotesize Upper bound, $\murx + 2\mutx > 2.2446$}, {\footnotesize Inner bound, $\murx + 2\mutx > 2.2446$}, {\footnotesize Upper bound, $\murx = 0.2, \mutx = 0.1$},{\footnotesize  Inner bound, $\murx = 0.2, \mutx = 0.1$}}
    
  
\end{axis}
\end{tikzpicture}
%
%
%
%
%
\caption{Bounds on region $\mathcal{S}^\star(\mu_{\Tx},\mu_{\Rx},\D)$ for $\D = 20$, $M = 3$, $ t = 4$ and for large and small conferencing rates.}
\label{fig6}
\vspace*{-3ex}
\end{figure}
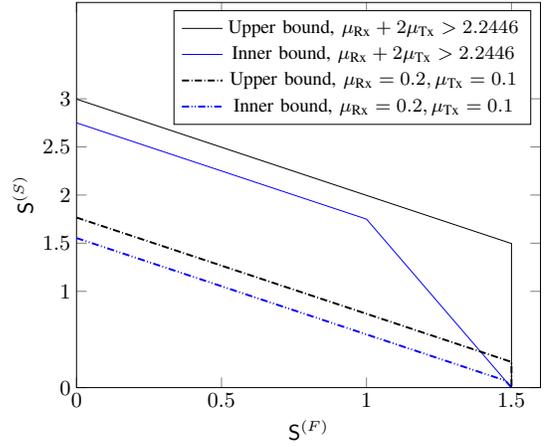

\begin{theorem} \label{lemma2}
The multiplexing gain  region satisfies
 \begin{IEEEeqnarray}{rCl} 
\S^{(F)} &  \le &\frac{M}{2} \label{upphex1}\\
 \lefteqn{\hspace{-.7cm}\S^{(F)}+\S^{(S)} \hspace{10cm}} \nonumber \\
 &&\hspace{-0.75cm} \le \min \Bigg \{ \frac{M}{2} + \frac{2\mu_{\Rx}  + 4\mu_{\Tx}}{3},  M \Bigg(1- \frac{1}{2(1+\D+\D^2)}\Bigg) \Bigg \}.\nonumber \\\label{upphex2} 
 \end{IEEEeqnarray}
 \end{theorem}
\begin{IEEEproof}
  See  Section \ref{hex:converse:proof}.
 \end{IEEEproof}
Inner and outer bounds on the multiplexing gain region are illustrated in Figure \ref{fig6}. For small cooperation prelogs, even when the  ``fast" multiplexing gain $\S^{(F)}$ is large, the sum multiplexing gain is almost the same as when only ``slow" messages are sent. For moderate or large cooperation prelogs, the sum multiplexing gain of our scheme remains constant over a wide range of  $\S^{(F)}$ where  the  stringent delay constraint  on ``fast" messages thus does  not harm the overall performance. For large ``fast" multiplexing gains, increasing it further by $\Delta>0$ seems to decrease the maximum ``slow" multiplexing gain by $4 \Delta$ and thus the sum multiplexing gain by $3 \Delta$. So, for any extra prelog on ``fast" multiplexing gain, one will have to sacrifice three times of that on the overall performance (sum multiplexing gain).



 \section{Proof of Theorem \ref{lemma1}} \label{sec:ach.th1}
 To achieve the performance in Theorem~\ref{lemma1},  Schemes 1--5 explained in the following need to be time-shared depending on the operating point and on the available conferencing prelogs. Notice that Schemes 
 2 and 3 are dual (i.e., achieve the same performance with same sum conferencing-prelog) when $t \le \frac{D}{4}$ and Schemes 4 and 5 are dual   when  $t \le \frac{D-2}{4}$.  By timesharing Schemes 2 and 3 and Schemes 4 and 5 in appropriate ratios, one can thus achieve the multiplexing-gain pairs in \eqref{pair3}, for $t \in \{1,\ldots, \frac{D}{4}\}$, and in \eqref{pair4}, for $t \in \{1,\ldots, \frac{D-2}{4} \}$, whenever   the sum of the cooperation-prelogs is sufficiently large. For setups where the  sum of the cooperation-prelogs is not sufficiently large, the schemes need to be further time-shared with the no-cooperation Scheme 1.  This establishes achievability of \eqref{s1} and \eqref{s3}.  Achievability of \eqref{s2}  is established by time-sharing Scheme 2, for  $t \in \{ \frac{D}{4}+1,\ldots, \frac{D-2}{2}\}$, or Scheme 5, for \eqref{s4}  with the no-cooperation Scheme 1. 
%

\par \underline{Scheme 1 \cite{Jafar}:} (\emph{Transmitting only ``fast" messages})
Transmission is based on \emph{interference alignment} \cite{Jafar}, and cooperation links are completely ignored. This scheme achieves multiplexing gain pair $\big( \S^{(F)}= \frac{M}{2} , \ \S^{(S)} = 0\big)$ for any conferencing prelogs  $\mu_{\text{Tx},1},\mu_{\text{Rx},1} \geq  0$.
  \par \underline{Scheme 2 \cite{Samet}:} (\emph{Transmitting only ``slow" messages with only Rx conferencing}). For details, see \cite{Samet}. The scheme has parameter $t\in\{1, \ldots, \D/2\}$ and achieves 
  multiplexing gain pair $\big( \S^{(F)}= 0 , \; \S^{(S)} = M\frac{3t-1}{3t}\big)$ using cooperation prelogs $\mu_{\text{Tx},2} = 0$ and $\mu_{\text{Rx},2} = M \frac{2t-1}{3}$.
  
  \par
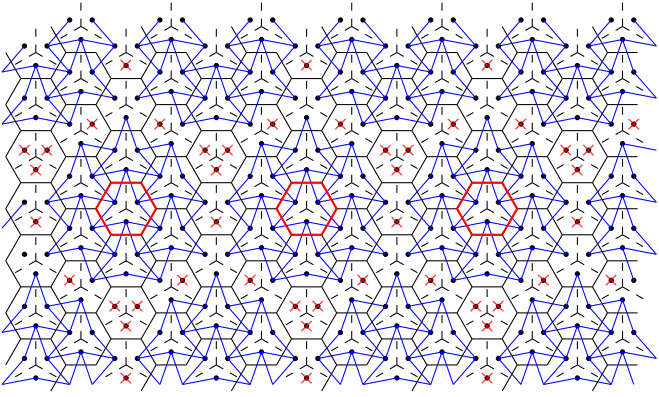
\begin{figure}[t]
  	\centering
  	\footnotesize
  	\begin{tikzpicture}[scale=0.4, >=stealth]
  	\centering
  	\foreach \i in {0,...,6} 
  	\foreach \j in {1,...,6} {
  		\foreach \a in {0,120,-120} \draw[black] (3*\i,2*sin{60}*\j) -- +(\a:1);}
  	\foreach \i in {0,...,6} 
  	\foreach \j in {0,...,6} {
  		\foreach \a in {0,120,-120} \draw [black](3*\i+3*cos{60},2*sin{60}*\j+sin{60}) -- +(\a:1);}
  	\foreach \ii in {0,...,6}
  	\foreach \jj in {0,...,6}{
  		\foreach \a in {90,-30,-150} \draw[black,dashed] (0.5+3*\jj,0.8660+\ii*1.7321) -- +(\a:0.8660);
  		\foreach \a in {90,-30,-150} \draw[black,dashed] (2+3*\jj,1.7321+\ii*1.7321) -- +(\a:0.8660);}
  	\foreach \c in {0,...,6}
  	\foreach \cc in {0,...,6}{
  		\draw [fill=black] (0.5 + 3*\c, 1.7321/4+1.7321*\cc) circle (0.07);}
  	\foreach \c in {0,...,6}
  	\foreach \cc in {0,...,6}{
  		\draw [fill=black] (2+ 3*\c, 3*1.7321/4+1.7321*\cc) circle (0.07);}
  	\foreach \ccc in {0,...,6}
  	\foreach \cccc in {0,...,6}{
  		\draw [fill=black] (0.5+0.5*0.75+ 3*\ccc, 1.7321/2+1.7321/8+1.7321*\cccc) circle (0.07);
  		\draw [fill=black] (2-0.5*0.75+3*\ccc, 1.7321/2+1.7321/2+1.7321/8+1.7321*\cccc) circle (0.07);
  		\draw [fill=black] (0.5-0.5*0.75+ 3*\ccc, 1.7321/2+1.7321/8+1.7321*\cccc) circle (0.07);
  		\draw [fill=black] (2+0.5*0.75+3*\ccc, 1.7321/2+1.7321/2+1.7321/8+1.7321*\cccc) circle (0.07);}
  	
  	\foreach \c in {0,2,4,6}
  	\foreach \cc in {3,4}{
  		\node[draw =none, rotate =45] (s2) at (0.5 + 3*\c, 1.7321/4+1.7321*\cc) {{\color{red}$+$}};}
  	
  	\foreach \c in {1,3,5}
  	\foreach \cc in {0,1,6}{
  		\node[draw =none, rotate =45] (s2) at (0.5 + 3*\c, 1.7321/4+1.7321*\cc) {{\color{red}$+$}};}
  	
  	\foreach \ccc in {0,2,4,6}
  	\foreach \cccc in {4}{
  		\node[draw =none, rotate =45] (s2) at (0.5+0.5*0.75+ 3*\ccc, 1.7321/2+1.7321/8+1.7321*\cccc) {{\color{red}$+$}};}
  	
  	\foreach \ccc in {1,3,5}
  	\foreach \cccc in {1}{
  		\node[draw =none, rotate =45] (s2) at (0.5+0.5*0.75+ 3*\ccc, 1.7321/2+1.7321/8+1.7321*\cccc) {{\color{red}$+$}};}
  	
  	\foreach \ccc in {0,2,4,6}
  	\foreach \cccc in {1}{
  		\node[draw =none, rotate =45] (s2) at (2-0.5*0.75+3*\ccc, 1.7321/2+1.7321/2+1.7321/8+1.7321*\cccc) {{\color{red}$+$}};}
  	
  	\foreach \ccc in {1,3,5}
  	\foreach \cccc in {4}{
  		\node[draw =none, rotate =45] (s2) at (2-0.5*0.75+3*\ccc, 1.7321/2+1.7321/2+1.7321/8+1.7321*\cccc) {{\color{red}$+$}};}
  	
  	\foreach \ccc in {0,2,4,6}
  	\foreach \cccc in {4}{
  		\node[draw =none, rotate =45] (s2) at (0.5-0.5*0.75+ 3*\ccc, 1.7321/2+1.7321/8+1.7321*\cccc) {{\color{red}$+$}};}
  	
  	\foreach \ccc in {1,3,5}
  	\foreach \cccc in {1}{
  		\node[draw =none, rotate =45] (s2) at (0.5-0.5*0.75+ 3*\ccc, 1.7321/2+1.7321/8+1.7321*\cccc) {{\color{red}$+$}};}
  	
  	\foreach \ccc in {0,2,4,6}
  	\foreach \cccc in {4}{
  		\node[draw =none, rotate =45] (s2) at (2+0.5*0.75+3*\ccc, 1.7321/2+1.7321/2+1.7321/8+1.7321*\cccc) {{\color{red}$+$}};}
  	\foreach \ccc in {1,3,5}
  	\foreach \cccc in {1}{
  		\node[draw =none, rotate =45] (s2) at (2+0.5*0.75+3*\ccc, 1.7321/2+1.7321/2+1.7321/8+1.7321*\cccc) {{\color{red}$+$}};}
  	\foreach \cc in {1,2,6}
  	\foreach \c in {0,2,4,6}{
  		\draw [blue] (0.5+3*\c , 1.7321/4+1.7321*\cc) --(0.5+0.5*0.75+ 3*\c, 1.7321/2+1.7321/8+1.7321*\cc- 1.7321);
  		\draw [blue] (0.5+3*\c , 1.7321/4+1.7321*\cc) --(0.5-0.5*0.75+ 3*\c, 1.7321/2+1.7321/8+1.7321*\cc- 1.7321);}
  	\foreach \cc in {3,4,5}
  	\foreach \c in {1,3,5}{
  		\draw [blue] (0.5+3*\c , 1.7321/4+1.7321*\cc) --(0.5+0.5*0.75+ 3*\c, 1.7321/2+1.7321/8+1.7321*\cc- 1.7321);
  		\draw [blue] (0.5+3*\c , 1.7321/4+1.7321*\cc) --(0.5-0.5*0.75+ 3*\c, 1.7321/2+1.7321/8+1.7321*\cc- 1.7321);}
  	\foreach \cc in {0,1,5,6}
  	\foreach \c in {0,2,4,6}{
  		\draw [blue] (0.5+3*\c , 1.7321/4+1.7321*\cc) --(2-0.5*0.75+3*\c, 1.7321/2+1.7321/2+1.7321/8+1.7321*\cc- 1.7321);
  		\draw [blue] (0.5+3*\c , 1.7321/4+1.7321*\cc) --(2+0.5*0.75+3*\c-3, 1.7321/2+1.7321/2+1.7321/8+1.7321*\cc- 1.7321);}
  	
  	\foreach \cc in {2,3,4}
  	\foreach \c in {1,3,5}{
  		\draw [blue] (0.5+3*\c , 1.7321/4+1.7321*\cc) --(2-0.5*0.75+3*\c, 1.7321/2+1.7321/2+1.7321/8+1.7321*\cc- 1.7321);
  		\draw [blue] (0.5+3*\c , 1.7321/4+1.7321*\cc) --(2+0.5*0.75+3*\c-3, 1.7321/2+1.7321/2+1.7321/8+1.7321*\cc- 1.7321);}
  	
  	\foreach \cc in {0,1,2,3,5,6}
  	\foreach \c in {0,2,4,6}{
  		\draw [blue] (2+3*\c , 3*1.7321/4+1.7321*\cc) --(0.5+0.5*0.75+ 3*\c, 1.7321/2+1.7321/8+1.7321*\cc);}
  	\foreach \cc in {0,2,3,4,5,6}
  	\foreach \c in {1,3,5}{
  		\draw [blue] (2+3*\c , 3*1.7321/4+1.7321*\cc) --(0.5+0.5*0.75+ 3*\c, 1.7321/2+1.7321/8+1.7321*\cc);} 
  	\foreach \cc in {0,2,3,5,6}
  	\foreach \c in {0,...,6}{
  		\draw [blue] (0.5+0.5*0.75+3*\c, 1.7321/2+1.7321/8+1.7321*\cc) -- (2-0.5*0.75+ 3*\c, 1.7321/2+1.7321/2+1.7321/8+1.7321*\cc);}
  	
  	\foreach \cc in {0,2,3,4,5,6}
  	\foreach \c in {0,2,4,6}{
  		\draw [blue] (2+3*\c , 3*1.7321/4+1.7321*\cc) --(0.5-0.5*0.75+ 3*\c+3, 1.7321/2+1.7321/8+1.7321*\cc);}
  	
  	\foreach \cc in {0,1,2,3,5,6}
  	\foreach \c in {1,3,5}{
  		\draw [blue] (2+3*\c , 3*1.7321/4+1.7321*\cc) --(0.5-0.5*0.75+ 3*\c+3, 1.7321/2+1.7321/8+1.7321*\cc);}
  	\foreach \cc in {0,1,3,4,5,6}
  	\foreach \c in {0,2,4,6}{
  		\draw [blue] (2+3*\c , 3*1.7321/4+1.7321*\cc) --(2-0.5*0.75+3*\c, 1.7321/2+1.7321/2+1.7321/8+1.7321*\cc- 1.7321);}
  	
  	\foreach \cc in {0,2,1,3,4,6}
  	\foreach \c in {1,3,5}{
  		\draw [blue] (2+3*\c , 3*1.7321/4+1.7321*\cc) --(2-0.5*0.75+3*\c, 1.7321/2+1.7321/2+1.7321/8+1.7321*\cc- 1.7321);}
  	
  	\foreach \cc in {0,1,2,3,4,6}
  	\foreach \c in {0,2,4,6}{
  		\draw [blue] (2+3*\c , 3*1.7321/4+1.7321*\cc) --(2+0.5*0.75+3*\c, 1.7321/2+1.7321/2+1.7321/8+1.7321*\cc- 1.7321);}
  	
  	\foreach \cc in {0,1,3,4,5,6}
  	\foreach \c in {1,3,5}{
  		\draw [blue] (2+3*\c , 3*1.7321/4+1.7321*\cc) --(2+0.5*0.75+3*\c, 1.7321/2+1.7321/2+1.7321/8+1.7321*\cc- 1.7321);}
  	
  	\foreach \cc in {0,2,3,4,6}
  	\foreach \c in {1,3,5}{   
  		\draw [blue] (0.5+0.5*0.75+3*\c, 1.7321/2+1.7321/8+1.7321*\cc) -- (2-0.5*0.75+ 3*\c, 1.7321/2+1.7321/2+1.7321/8+1.7321*\cc- 1.7321);}
  	
  	\foreach \cc in {0,1,3,5,6}
  	\foreach \c in {0,2,4,6}{   
  		\draw [blue] (0.5+0.5*0.75+3*\c, 1.7321/2+1.7321/8+1.7321*\cc) -- (2-0.5*0.75+ 3*\c, 1.7321/2+1.7321/2+1.7321/8+1.7321*\cc- 1.7321);}
  	
  	\foreach \cc in {0,2,3,5,6}
  	\foreach \c in {0,...,6}{
  		\draw [blue] (2+0.5*0.75+3*\c, 1.7321/2+1.7321/2+1.7321/8+1.7321*\cc) -- (0.5-0.5*0.75+ 3+3*\c, 1.7321/2+1.7321/8+1.7321*\cc);}
  	
  	\foreach \cc in {-1,1,2,3,5}
  	\foreach \c in {0,2,4}{
  		\draw [blue] (2+0.5*0.75+3*\c, 1.7321/2+1.7321/2+1.7321/8+1.7321*\cc) -- (0.5-0.5*0.75+ 3+3*\c, 1.7321/2+1.7321/8+1.7321+1.7321*\cc);}
  	\foreach \cc in {-1,0,2,4,5}
  	\foreach \c in {-1,1,3,5}{
  		\draw [blue] (2+0.5*0.75+3*\c, 1.7321/2+1.7321/2+1.7321/8+1.7321*\cc) -- (0.5-0.5*0.75+ 3+3*\c, 1.7321/2+1.7321/8+1.7321+1.7321*\cc);}
  	
  	\foreach \a in {0,120} \draw[red,thick] (3*3,2*sin{60}*3) -- +(\a:1);
  	\foreach \a in {0,-120} \draw[red,thick] (3*3,2*sin{60}*4) -- +(\a:1);
  	\foreach \a in {120,-120} \draw [red,thick](3*3+3*cos{60},2*sin{60}*3+sin{60}) -- +(\a:1);
  	
  	\foreach \a in {0,120} \draw[red,thick] (3,2*sin{60}*3) -- +(\a:1);
  	\foreach \a in {0,-120} \draw[red,thick] (3,2*sin{60}*4) -- +(\a:1);
  	\foreach \a in {120,-120} \draw [red,thick](3+3*cos{60},2*sin{60}*3+sin{60}) -- +(\a:1);
  	
  	\foreach \a in {0,120} \draw[red,thick] (3*5,2*sin{60}*3) -- +(\a:1);
  	\foreach \a in {0,-120} \draw[red,thick] (3*5,2*sin{60}*4) -- +(\a:1);
  	\foreach \a in {120,-120} \draw [red,thick](3*5+3*cos{60},2*sin{60}*3+sin{60}) -- +(\a:1);
  	
  	
  	\end{tikzpicture} 
  	\caption{Illustration of the non-interfering clusters for $t=2$. Red cells depict master cells and silenced mobile users are marked by the red multiply  icons.}
  	\label{fig2}
  	\vspace*{-3ex}
  \end{figure}

 \underline{Scheme 3:} (\emph{Transmitting only ``slow" messages with only Tx conferencing})
  Pick a parameter $t\in\{1,\ldots, \D/4\}$.   
 Define \textit{master cells} \cite{Samet} so that they build a regular grid of equilateral triangles where the three master cells forming each of the triangles lay $3t$ cell-hops apart from each other. (See Fig.~\ref{fig2}, where master cells are in red.) We silence a subset  of the mobile users lying in cells that are $t$ hops apart  from the closest master cell, so as to decompose the network into non-interfering clusters. This  avoids interference 
 propagating too far.  (See for example  Fig.~\ref{fig2}, for an example with $t=2$.) 
 Each non-silenced mobile user intends to send  a ``slow"  message, which it conveys to a dedicated  \emph{master user} in the closest master cell. Each master user employs standard Gaussian codebooks  to encode all the ``slow" messages. After precoding (which is explained in more detail later), it sends quantized  versions of the signals  over the cooperation links back to the mobile users in the same cluster. These mobile users reconstruct the quantized signals and  transmit them over the network to the BSs.  The precoding is chosen in a way that the signal observed at each BS, in each sector, only depends on the message sent in that sector and not on the messages sent in other sectors. Each message can thus be sent at multiplexing gain 1 and 
 the scheme  achieves  multiplexing gain pair 
 \begin{IEEEeqnarray}{rCl}\label{pair3}\big( \S^{(F)}= 0 , \; \S^{(S)} = M\frac{3t-1}{3t}\big).
 	\end{IEEEeqnarray} It requires at least  cooperation prelogs $\mu_{\text{Tx},3} = M \frac{2t-1}{3}$ and $\mu_{\text{Rx},3} = 0$.
 
  ~~

 \par \underline{Scheme 4 :} (\emph{Alternating ``fast" and ``slow" messages with more Tx conferencing})
 Pick a parameter $t\in\{1,\ldots, (\D-2)/4\}$. Define master cells and silence  mobile users as in scheme 3, see Figure~\ref{fig2}.  Each of the  remaining non-silenced users sends either a ``slow" or a ``fast" message. The proposed message assignment is  shown in Figure \ref{fig3}, where sectors with ``fast'' messages are  in green and  sectors with ``slow'' messages in red. The idea is to pack as many users with ``fast" messages under the constraint that sectors with ``fast" messages do not interfere. Therefore, the way we assigned ``slow" and ``fast" messages, communication of ``fast'' messages is interfered only by ``slow'' messages. 

In each master cell, consider one dedicated mobile user, called \emph{master user}, that will coordinate the transmission.
In fact, each mobile user sends its ``slow'' message to the closest master user. The master user encodes all received ``slow" messages using individual Gaussian codebooks and  then  precodes the Gaussian codewords in a way that when the precoded streams are transmitted over \emph{all} the active antennas in the cluster, then the signal observed in each sector only depends on the ``slow" message sent in that sector. (Since some active mobile users do not send  any ``slow" message at all, this means that the precoding ensures that the transmitted signals are nulled out in these sectors.) The master user finally applies a Gaussian vector quantizer on each precoded antenna and sends the corresponding quantization information over the cooperation links  to the corresponding mobile user. Each of these mobile users lies in the same cluster as the master user, and therefore this communication takes less than $2t < \frac{\D}{2}$ cooperation rounds. All active  mobile users then  reconstruct the quantized signals intended for them. Users sending ``slow" messages simply send this reconstructed signal over the network to the BSs. Users sending ``fast" messages also encode their ``fast" message using a Gaussian codebook and transmit  the sum of this codeword with the reconstructed signal over the network. 

 The BSs in the green sectors can decode their ``fast'' messages based on almost (up to quantization errors) interference-free signals since communication of ``fast'' messages is interfered only by ``slow'' messages, whose interference has been canceled in the precoding.  After decoding, these BSs send their decoded  ``fast'' messages over  the cooperation links  to the neighbouring receivers, which then cancel  interference from their received signals. Thanks to the applied precoding, the resulting signals  at a given BS only depends (up to some quantization errors up to noise level)  on the ``slow" message intended to this BS, without any interference from  other ``slow" or ``fast" messages. 
 
 In the described scheme, each transmitted message can be sent at multiplexing gain 1. The fraction of users in the network that are inactive is $\frac{1}{3t}$. The fraction of   users sending ``slow'' messages  is  $\frac{(2t -1)}{3t}$ and   the fraction of users sending ``fast'' messages  is $\frac{1}{3}$. The scheme thus achieves the multiplexing-gain pair
 \begin{equation} \label{pair4}
 \Big( \S^{(F)}= \frac{M}{3} , \; \S^{(S)} = M\frac{2t-1}{3t}\Big).
 \end{equation}
 Notice that Scheme 4 achieves the same sum multiplexing gain as Schemes 2 and  3, which send  only ``slow" messages.

 The described scheme requires $\Dt \ge 4t + 1$ Tx-conferencing rounds and  one single Rx-conferencing rounds.
 The number of Tx-conferencing messages (each of  prelog  1) sent in a given cluster is $2Mt(8t^2 + 3t -2)$. Since there are $12\cdot3t^2$ cooperation links  in each cluster,  the scheme requires a 
Tx-conferencing prelog of at least
\begin{equation}\label{mutx1}
\mu_{\Tx,4}  = \frac{ 2Mt(8t^2 +3t -2)}{12\cdot3t^2}.
\end{equation}
The number of Rx-conferencing messages (each of prelog 1) sent in a given cluster  is $3M(3t^2 -1)$. 
Since there are $6\cdot3t^2$ Rx-conferencing links in each cluster, the scheme requires a Rx-cooperation prelog of at least
\begin{equation}\label{murx1}
\mu_{\Rx, 4}  = \frac{3M(3t^2 -1)}{6\cdot3t^2}.
\end{equation}

 \begin{figure}[t]
  \centering
\begin{tikzpicture}[scale=0.45, >=stealth]
  \tikzset{
    box/.style={x=11.5mm,y=6.4mm,regular polygon,regular polygon sides=6, minimum size=15mm, inner sep=0mm, outer sep=0mm, rotate=0, draw },
      box2/.style={x=49.2mm,y=32.2mm,regular polygon,regular polygon sides=6, minimum size=104mm, inner sep=0mm, outer sep=0mm, rotate=0, draw, green, rotate = 90 },
    filledhex/.style={box,minimum size=15mm,draw=none, fill=green!30}
  }

\fillaroundd{0}{5}

\foreach \i in {0,...,3}  {
\draw [fill = red!40] (5.75-0.4 ,5.75+0.65+ 1.28*\i)--(5.175,5.75+0.965+ 1.28*\i)-- (5.75,5.75+ 1.28+ 1.28*\i)--(6.325,5.75+0.965+ 1.28*\i) --(5.75+0.4,5.75+0.65+ 1.28*\i)-- (5.75-0.4,5.75+0.65+ 1.28*\i);}
\foreach \i in {0,...,2}  {
\draw [fill = green!30] (5.175,5.75+0.965+ 1.28*\i)-- (5.75,5.75+ 1.28+ 1.28*\i)--(5.75,5.75+ 1.92+ 1.28*\i) --(5.75-0.4,5.75+ 1.92+ 1.28*\i)-- (5,5.75+ 1.28+ 1.28*\i)--(5.175,5.75+0.965+ 1.28*\i);
\draw [fill = green!30] (6.325,5.75+0.965+ 1.28*\i)-- (5.75,5.75+ 1.28+ 1.28*\i)--(5.75,5.75+ 1.92+ 1.28*\i) --(5.75+0.4,5.75+ 1.92+ 1.28*\i)-- (6.5,5.75+ 1.28+ 1.28*\i)--(6.325,5.75+0.965+ 1.28*\i);}
\foreach \i in {-2,-3,-4}  {
\draw [fill = green!30] (5.75-0.4 ,5.75+0.65+ 1.28*\i)--(5.175,5.75+0.965+ 1.28*\i)-- (5.75,5.75+ 1.28+ 1.28*\i)--(6.325,5.75+0.965+ 1.28*\i) --(5.75+0.4,5.75+0.65+ 1.28*\i)-- (5.75-0.4,5.75+0.65+ 1.28*\i);}
\foreach \i in {-2,-3,-4}  {
\draw [fill = red!40] (5.175,5.75+0.965+ 1.28*\i)-- (5.75,5.75+ 1.28+ 1.28*\i)--(5.75,5.75+ 1.92+ 1.28*\i) --(5.75-0.4,5.75+ 1.92+ 1.28*\i)-- (5,5.75+ 1.28+ 1.28*\i)--(5.175,5.75+0.965+ 1.28*\i);
\draw [fill = red!40] (6.325,5.75+0.965+ 1.28*\i)-- (5.75,5.75+ 1.28+ 1.28*\i)--(5.75,5.75+ 1.92+ 1.28*\i) --(5.75+0.4,5.75+ 1.92+ 1.28*\i)-- (6.5,5.75+ 1.28+ 1.28*\i)--(6.325,5.75+0.965+ 1.28*\i);}
\foreach \i in {0,...,3}{
\draw [fill = red!40] (5.75-0.4 - 1.15 ,5.75+0.65+ 1.28*\i-0.64)--(5.175- 1.15,5.75+0.965+ 1.28*\i-0.64)-- (5.75- 1.15,5.75+ 1.28+ 1.28*\i-0.64)--(6.325- 1.15,5.75+0.965+ 1.28*\i-0.64) --(5.75- 1.15+0.4,5.75+0.65+ 1.28*\i-0.64)-- (5.75-0.4- 1.15,5.75+0.65+ 1.28*\i-0.64);}
\foreach \i in {-1,...,2}{
\draw [fill = green!30] (5.175 - 1.15 ,5.75+0.965+ 1.28*\i-0.64)-- (5.75 - 1.15 ,5.75+ 1.28+ 1.28*\i-0.64)--(5.75 - 1.15 ,5.75+ 1.92+ 1.28*\i-0.64) --(5.75-0.4 - 1.15 ,5.75+ 1.92+ 1.28*\i-0.64)-- (5 - 1.15 ,5.75+ 1.28+ 1.28*\i-0.64)--(5.175 - 1.15 ,5.75+0.965+ 1.28*\i-0.64);
\draw [fill = red!40] (6.325 - 1.15,5.75+0.965+ 1.28*\i-0.64)-- (5.75 - 1.15,5.75+ 1.28+ 1.28*\i-0.64)--(5.75 - 1.15,5.75+ 1.92+ 1.28*\i-0.64) --(5.75+0.4 - 1.15,5.75+ 1.92+ 1.28*\i-0.64)-- (6.5 - 1.15,5.75+ 1.28+ 1.28*\i-0.64)--(6.325 - 1.15,5.75+0.965+ 1.28*\i-0.64);}
\foreach \i in {-1,...,-3}{
\draw [fill = green!30] (5.75-0.4 - 1.15 ,5.75+0.65+ 1.28*\i-0.64)--(5.175- 1.15,5.75+0.965+ 1.28*\i-0.64)-- (5.75- 1.15,5.75+ 1.28+ 1.28*\i-0.64)--(6.325- 1.15,5.75+0.965+ 1.28*\i-0.64) --(5.75- 1.15+0.4,5.75+0.65+ 1.28*\i-0.64)-- (5.75-0.4- 1.15,5.75+0.65+ 1.28*\i-0.64);}
\foreach \i in {-2,...,-3}{
\draw [fill = red!40] (5.175 - 1.15 ,5.75+0.965+ 1.28*\i-0.64)-- (5.75 - 1.15 ,5.75+ 1.28+ 1.28*\i-0.64)--(5.75 - 1.15 ,5.75+ 1.92+ 1.28*\i-0.64) --(5.75-0.4 - 1.15 ,5.75+ 1.92+ 1.28*\i-0.64)-- (5 - 1.15 ,5.75+ 1.28+ 1.28*\i-0.64)--(5.175 - 1.15 ,5.75+0.965+ 1.28*\i-0.64);}
\foreach \i in {-1,...,-4}{
\draw [fill = red!40] (6.325 - 1.15,5.75+0.965+ 1.28*\i-0.64)-- (5.75 - 1.15,5.75+ 1.28+ 1.28*\i-0.64)--(5.75 - 1.15,5.75+ 1.92+ 1.28*\i-0.64) --(5.75+0.4 - 1.15,5.75+ 1.92+ 1.28*\i-0.64)-- (6.5 - 1.15,5.75+ 1.28+ 1.28*\i-0.64)--(6.325 - 1.15,5.75+0.965+ 1.28*\i-0.64);}
\foreach \i in {0,...,3}{
\draw [fill = red!40] (5.75-0.4 +1.15 ,5.75+0.65+ 1.28*\i-0.64)--(5.175+ 1.15,5.75+0.965+ 1.28*\i-0.64)-- (5.75+ 1.15,5.75+ 1.28+ 1.28*\i-0.64)--(6.325+1.15,5.75+0.965+ 1.28*\i-0.64) --(5.75+1.15+0.4,5.75+0.65+ 1.28*\i-0.64)-- (5.75-0.4+ 1.15,5.75+0.65+ 1.28*\i-0.64);}
\foreach \i in {-1,...,2}{
\draw [fill = red!40] (5.175 +1.15 ,5.75+0.965+ 1.28*\i-0.64)-- (5.75 +1.15 ,5.75+ 1.28+ 1.28*\i-0.64)--(5.75 +1.15 ,5.75+ 1.92+ 1.28*\i-0.64) --(5.75-0.4 + 1.15 ,5.75+ 1.92+ 1.28*\i-0.64)-- (5 +1.15 ,5.75+ 1.28+ 1.28*\i-0.64)--(5.175 + 1.15 ,5.75+0.965+ 1.28*\i-0.64);
\draw [fill = green!30] (6.325 +1.15,5.75+0.965+ 1.28*\i-0.64)-- (5.75 + 1.15,5.75+ 1.28+ 1.28*\i-0.64)--(5.75 + 1.15,5.75+ 1.92+ 1.28*\i-0.64) --(5.75+0.4 + 1.15,5.75+ 1.92+ 1.28*\i-0.64)-- (6.5 + 1.15,5.75+ 1.28+ 1.28*\i-0.64)--(6.325 + 1.15,5.75+0.965+ 1.28*\i-0.64);}
\foreach \i in {-1,...,-3}{
\draw [fill = green!30] (5.75-0.4 + 1.15 ,5.75+0.65+ 1.28*\i-0.64)--(5.175+ 1.15,5.75+0.965+ 1.28*\i-0.64)-- (5.75 + 1.15,5.75+ 1.28+ 1.28*\i-0.64)--(6.325 + 1.15,5.75+0.965+ 1.28*\i-0.64) --(5.75 +  1.15+0.4,5.75+0.65+ 1.28*\i-0.64)-- (5.75-0.4 + 1.15,5.75+0.65+ 1.28*\i-0.64);}
\foreach \i in {-1,...,-4}{
\draw [fill = red!40] (5.175 + 1.15 ,5.75+0.965+ 1.28*\i-0.64)-- (5.75 + 1.15 ,5.75+ 1.28+ 1.28*\i-0.64)--(5.75 + 1.15 ,5.75+ 1.92+ 1.28*\i-0.64) --(5.75-0.4 + 1.15 ,5.75+ 1.92+ 1.28*\i-0.64)-- (5 +1.15 ,5.75+ 1.28+ 1.28*\i-0.64)--(5.175 +1.15 ,5.75+0.965+ 1.28*\i-0.64);}
\foreach \i in {-2,...,-3}{
\draw [fill = red!40] (6.325 + 1.15,5.75+0.965+ 1.28*\i-0.64)-- (5.75 + 1.15,5.75+ 1.28+ 1.28*\i-0.64)--(5.75 + 1.15,5.75+ 1.92+ 1.28*\i-0.64) --(5.75+0.4 +1.15,5.75+ 1.92+ 1.28*\i-0.64)-- (6.5 +1.15,5.75+ 1.28+ 1.28*\i-0.64)--(6.325 + 1.15,5.75+0.965+ 1.28*\i-0.64);}
\foreach \j in {-1,1}
\foreach \i in {-1,...,2}  {
\draw [fill = red!40] (5.75-0.4+ 2.3*\j ,5.75+0.65+ 1.28*\i)--(5.175+ 2.3*\j,5.75+0.965+ 1.28*\i)-- (5.75+ 2.3*\j,5.75+ 1.28+ 1.28*\i)--(6.325+ 2.3*\j,5.75+0.965+ 1.28*\i) --(5.75+0.4+ 2.3*\j,5.75+0.65+ 1.28*\i)-- (5.75-0.4+ 2.3*\j,5.75+0.65+ 1.28*\i);}
\foreach \i in {-2,-3}  {
\draw [fill = green!30] (5.75-0.4+ 2.3 ,5.75+0.65+ 1.28*\i)--(5.175+ 2.3,5.75+0.965+ 1.28*\i)-- (5.75+ 2.3,5.75+ 1.28+ 1.28*\i)--(6.325+ 2.3,5.75+0.965+ 1.28*\i) --(5.75+0.4+ 2.3,5.75+0.65+ 1.28*\i)-- (5.75-0.4+ 2.3,5.75+0.65+ 1.28*\i);}

\foreach \i in {-4,...,1}  {
\draw [fill = red!40] (5.175+ 2.3,5.75+0.965+ 1.28*\i)-- (5.75+ 2.3,5.75+ 1.28+ 1.28*\i)--(5.75+ 2.3,5.75+ 1.92+ 1.28*\i) --(5.75-0.4+ 2.3,5.75+ 1.92+ 1.28*\i)-- (5+ 2.3,5.75+ 1.28+ 1.28*\i)--(5.175+ 2.3,5.75+0.965+ 1.28*\i);}
\foreach \i in {-2,...,1}  {
\draw [fill = green!30] (6.325+ 2.3,5.75+0.965+ 1.28*\i)-- (5.75+ 2.3,5.75+ 1.28+ 1.28*\i)--(5.75+ 2.3,5.75+ 1.92+ 1.28*\i) --(5.75+ 2.3+0.4,5.75+ 1.92+ 1.28*\i)-- (6.5+ 2.3,5.75+ 1.28+ 1.28*\i)--(6.325+ 2.3,5.75+0.965+ 1.28*\i);}
\foreach \i in {-3}  {
\draw [fill = red!40] (6.325+ 2.3,5.75+0.965+ 1.28*\i)-- (5.75+ 2.3,5.75+ 1.28+ 1.28*\i)--(5.75+ 2.3,5.75+ 1.92+ 1.28*\i) --(5.75+ 2.3+0.4,5.75+ 1.92+ 1.28*\i)-- (6.5+ 2.3,5.75+ 1.28+ 1.28*\i)--(6.325+ 2.3,5.75+0.965+ 1.28*\i);}
\foreach \i in {-2,-3}  {
\draw [fill = green!30] (5.75-0.4 -2.3,5.75+0.65+ 1.28*\i)--(5.175-2.3,5.75+0.965+ 1.28*\i)-- (5.75-2.3,5.75+ 1.28+ 1.28*\i)--(6.325-2.3,5.75+0.965+ 1.28*\i) --(5.75-2.3+0.4,5.75+0.65+ 1.28*\i)-- (5.75-0.4-2.3,5.75+0.65+ 1.28*\i);}
\foreach \i in {-3}  {
\draw [fill = red!40] (5.175-2.3,5.75+0.965+ 1.28*\i)-- (5.75-2.3,5.75+ 1.28+ 1.28*\i)--(5.75-2.3,5.75+ 1.92+ 1.28*\i) --(5.75-0.4-2.3,5.75+ 1.92+ 1.28*\i)-- (5-2.3,5.75+ 1.28+ 1.28*\i)--(5.175-2.3,5.75+0.965+ 1.28*\i);}
\foreach \i in {-2,-1,0,1}  {
\draw [fill = green!30] (5.175-2.3,5.75+0.965+ 1.28*\i)-- (5.75-2.3,5.75+ 1.28+ 1.28*\i)--(5.75-2.3,5.75+ 1.92+ 1.28*\i) --(5.75-0.4-2.3,5.75+ 1.92+ 1.28*\i)-- (5-2.3,5.75+ 1.28+ 1.28*\i)--(5.175-2.3,5.75+0.965+ 1.28*\i);}
\foreach \i in {1,...,-4}  {
\draw [fill = red!40] (6.325-2.3,5.75+0.965+ 1.28*\i)-- (5.75-2.3,5.75+ 1.28+ 1.28*\i)--(5.75-2.3,5.75+ 1.92+ 1.28*\i) --(5.75-2.3+0.4,5.75+ 1.92+ 1.28*\i)-- (6.5-2.3,5.75+ 1.28+ 1.28*\i)--(6.325-2.3,5.75+0.965+ 1.28*\i);}
\foreach \i in {-1,...,2}{
\draw [fill = red!40] (5.75-0.4 - 1.15 -2.3,5.75+0.65+ 1.28*\i-0.64)--(5.175- 1.15-2.3,5.75+0.965+ 1.28*\i-0.64)-- (5.75- 1.15-2.3,5.75+ 1.28+ 1.28*\i-0.64)--(6.325- 1.15-2.3,5.75+0.965+ 1.28*\i-0.64) --(5.75- 1.15+0.4-2.3,5.75+0.65+ 1.28*\i-0.64)-- (5.75-0.4- 1.15-2.3,5.75+0.65+ 1.28*\i-0.64);}
\foreach \i in {-2}{
\draw [fill = green!30] (5.75-0.4 - 1.15 -2.3,5.75+0.65+ 1.28*\i-0.64)--(5.175- 1.15-2.3,5.75+0.965+ 1.28*\i-0.64)-- (5.75- 1.15-2.3,5.75+ 1.28+ 1.28*\i-0.64)--(6.325- 1.15-2.3,5.75+0.965+ 1.28*\i-0.64) --(5.75- 1.15+0.4-2.3,5.75+0.65+ 1.28*\i-0.64)-- (5.75-0.4- 1.15-2.3,5.75+0.65+ 1.28*\i-0.64);}
\foreach \i in {-2,...,1}{
\draw [fill = green!30] (5.175 - 1.15-2.3 ,5.75+0.965+ 1.28*\i-0.64)-- (5.75 - 1.15-2.3 ,5.75+ 1.28+ 1.28*\i-0.64)--(5.75 - 1.15-2.3 ,5.75+ 1.92+ 1.28*\i-0.64) --(5.75-0.4 - 1.15-2.3 ,5.75+ 1.92+ 1.28*\i-0.64)-- (5 - 1.15-2.3 ,5.75+ 1.28+ 1.28*\i-0.64)--(5.175 - 1.15-2.3 ,5.75+0.965+ 1.28*\i-0.64);}
\foreach \i in {-3,...,1}{
\draw [fill = red!40] (6.325 - 1.15-2.3,5.75+0.965+ 1.28*\i-0.64)-- (5.75 - 1.15-2.3,5.75+ 1.28+ 1.28*\i-0.64)--(5.75 - 1.15-2.3,5.75+ 1.92+ 1.28*\i-0.64) --(5.75+0.4 - 1.15-2.3,5.75+ 1.92+ 1.28*\i-0.64)-- (6.5 - 1.15-2.3,5.75+ 1.28+ 1.28*\i-0.64)--(6.325 - 1.15-2.3,5.75+0.965+ 1.28*\i-0.64);}
\foreach \i in {2,...,-1}{
\draw [fill = red!40] (5.75-0.4 + 1.15+2.3,5.75+0.65+ 1.28*\i-0.64)--(5.175+1.15+2.3,5.75+0.965+ 1.28*\i-0.64)-- (5.75+ 1.15+2.3,5.75+ 1.28+ 1.28*\i-0.64)--(6.325+ 1.15+2.3,5.75+0.965+ 1.28*\i-0.64) --(5.75+ 1.15+0.4+2.3,5.75+0.65+ 1.28*\i-0.64)-- (5.75-0.4+ 1.15+2.3,5.75+0.65+ 1.28*\i-0.64);}
\foreach \i in {-2}{
\draw [fill = green!30] (5.75-0.4 + 1.15+2.3,5.75+0.65+ 1.28*\i-0.64)--(5.175+1.15+2.3,5.75+0.965+ 1.28*\i-0.64)-- (5.75+ 1.15+2.3,5.75+ 1.28+ 1.28*\i-0.64)--(6.325+ 1.15+2.3,5.75+0.965+ 1.28*\i-0.64) --(5.75+ 1.15+0.4+2.3,5.75+0.65+ 1.28*\i-0.64)-- (5.75-0.4+ 1.15+2.3,5.75+0.65+ 1.28*\i-0.64);}
\foreach \i in {1,...,-3}{
\draw [fill = red!40] (5.175+ 1.15+2.3 ,5.75+0.965+ 1.28*\i-0.64)-- (5.75 + 1.15+2.3 ,5.75+ 1.28+ 1.28*\i-0.64)--(5.75+ 1.15+2.3 ,5.75+ 1.92+ 1.28*\i-0.64) --(5.75-0.4 + 1.15+2.3 ,5.75+ 1.92+ 1.28*\i-0.64)-- (5+ 1.15+2.3,5.75+ 1.28+ 1.28*\i-0.64)--(5.175 + 1.15+2.3,5.75+0.965+ 1.28*\i-0.64);}
\foreach \i in {1,...,-2}{
\draw [fill = green!30] (6.325 + 1.15+2.3,5.75+0.965+ 1.28*\i-0.64)-- (5.75 + 1.15+2.3,5.75+ 1.28+ 1.28*\i-0.64)--(5.75 + 1.15+2.3,5.75+ 1.92+ 1.28*\i-0.64) --(5.75+0.4 + 1.15+2.3,5.75+ 1.92+ 1.28*\i-0.64)-- (6.5 + 1.15+2.3,5.75+ 1.28+ 1.28*\i-0.64)--(6.325 + 1.15+2.3,5.75+0.965+ 1.28*\i-0.64);}
\foreach \i in {-3,...,0}  {
\draw [fill = red!40] (5.175+2*2.3,5.75+0.965+ 1.28*\i)-- (5.75+2*2.3,5.75+ 1.28+ 1.28*\i)--(5.75+2*2.3,5.75+ 1.92+ 1.28*\i) --(5.75-0.4+2*2.3,5.75+ 1.92+ 1.28*\i)-- (5+2*2.3,5.75+ 1.28+ 1.28*\i)--(5.175+2*2.3,5.75+0.965+ 1.28*\i);
\draw [fill = red!40] (6.325-4.6,5.75+0.965+ 1.28*\i)-- (5.75-4.6,5.75+ 1.28+ 1.28*\i)--(5.75-4.6,5.75+ 1.92+ 1.28*\i) --(5.75+0.4-4.6,5.75+ 1.92+ 1.28*\i)-- (6.5-4.6,5.75+ 1.28+ 1.28*\i)--(6.325-4.6,5.75+0.965+ 1.28*\i);}
\end{tikzpicture}
  \caption{Illustration of sector allocation in a cluster with $t=4$. Users in green sectors send ``fast'' messages and users in red sectors send ``slow'' messages.  }
  \label{fig3}
  \vspace*{-4ex}
\end{figure}
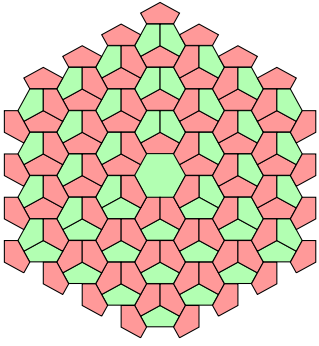~~%
\par \underline{Scheme 5 \cite{HomaITW2018}:} (\emph{Alternating ``fast" and ``slow" messages with more Rx conferencing}) The same  set of mobile users is silenced as in Scheme 4, but interference caused from ``slow" messages to other ``slow" messages is cancelled at the Rx side and not at the Tx side as in Scheme 4. (In contrast, interference cancellation  of ``slow" messages to ``fast" messages and of ``fast" messages to ``slow" messages is done as in Scheme 4: the former is canceled at the Tx side and the latter at the Rx side.) Another difference to Scheme 4 is that here the parameter $t$ can be chosen in the set  $\{1,\ldots, (\D-2)/2\}$. In fact, larger values of $t$ are possible here because less hops are required to communicate from one sector to another when communication is over BSs (there is only one for each triple of sectors) than when it is performed over mobile users (each sector has a different user). 
%
The scheme achieves the same multiplexing-gain pair as Scheme 4, i.e., \eqref{pair4}.
The required Tx-conferencing prelog is
\begin{equation}\label{mutx2}
\mu_{\Tx,5}  = \frac{ 6Mt(2t-1)}{12\cdot3t^2},
\end{equation}
and  the required Rx-conferencing prelog is 
\begin{equation}\label{murx2}
\mu_{\Rx,5}  = \frac{M(8t^3 + 6t^2 + t -3)}{6\cdot3t^2}
\end{equation}
\section{Proof of Theorem \ref{lemma2}} \label{hex:converse:proof}
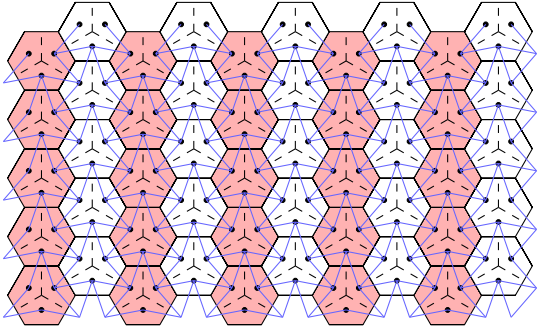
\begin{figure}[t]
  \centering
  \small
 \begin{tikzpicture}[scale=0.45, >=stealth]
 \centering
  \foreach \i in {-2,...,2} 
  \foreach \j in {0,...,4} {
\draw[ fill = red!30] (3*\i,2*sin{60}*\j+2*sin{60}) -- +(0:1) -- (3*\i+3*cos{60},2*sin{60}*\j+sin{60}) -- +(120:1) --(3*\i+3*cos{60},2*sin{60}*\j+sin{60}) -- +(-120:1)--(3*\i,2*sin{60}*\j) -- +(0:1)--(3*\i,2*sin{60}*\j) -- +(120:1)-- (3*\i,2*sin{60}*\j+2*sin{60}) -- +(-120:1)  ;}
  \foreach \i in {-2,...,2}
 \foreach \j in {0,...,4} {
 \draw[ fill= white] (3*\i+3*cos{60},2*sin{60}*\j+sin{60}) -- +(0:1)--(3*\i+3*cos{60},2*sin{60}*\j+sin{60}) -- +(120:1)--(3*\i+3*cos{60},2*sin{60}*\j+3*sin{60}) -- +(-120:1)--(3*\i+3*cos{60},2*sin{60}*\j+3*sin{60}) -- +(0:1)-- (3*\i+3,2*sin{60}*\j+2*sin{60}) -- +(120:1)--(3*\i +3,2*sin{60}*\j+2*sin{60}) -- +(-120:1);  }
  \foreach \ii in {0,...,4}
    \foreach \jj in {-2,...,2}{
  \foreach \a in {90,-30,-150} \draw[black,dashed] (0.5+3*\jj,0.8660+\ii*1.7321) -- +(\a:0.8660);
    \foreach \a in {90,-30,-150} \draw[black,dashed] (2+3*\jj,1.7321+\ii*1.7321) -- +(\a:0.8660);}
  \foreach \c in {-2,...,2}
     \foreach \cc in {0,...,4}{
 \draw [fill=black] (0.5 + 3*\c, 1.7321/4+1.7321*\cc) circle (0.07);}
  \foreach \c in {-2,...,2}
     \foreach \cc in {0,...,4}{
 \draw [fill=black] (2+ 3*\c, 3*1.7321/4+1.7321*\cc) circle (0.07);}
 \foreach \ccc in {-2,...,2}
     \foreach \cccc in {0,...,4}{
  \draw [fill=black] (0.5+0.5*0.75+ 3*\ccc, 1.7321/2+1.7321/8+1.7321*\cccc) circle (0.07);
   \draw [fill=black] (2-0.5*0.75+3*\ccc, 1.7321/2+1.7321/2+1.7321/8+1.7321*\cccc) circle (0.07);
     \draw [fill=black] (0.5-0.5*0.75+ 3*\ccc, 1.7321/2+1.7321/8+1.7321*\cccc) circle (0.07);
   \draw [fill=black] (2+0.5*0.75+3*\ccc, 1.7321/2+1.7321/2+1.7321/8+1.7321*\cccc) circle (0.07);}
    \foreach \cc in {1,...,4}
     \foreach \c in {-2,...,2}{
 \draw [blue!60] (0.5+3*\c , 1.7321/4+1.7321*\cc) --(0.5+0.5*0.75+ 3*\c, 1.7321/2+1.7321/8+1.7321*\cc- 1.7321);
 \draw [blue!60] (0.5+3*\c , 1.7321/4+1.7321*\cc) --(0.5-0.5*0.75+ 3*\c, 1.7321/2+1.7321/8+1.7321*\cc- 1.7321);}
 \foreach \cc in {0,...,4}
     \foreach \c in {-2,...,2}{
 \draw [blue!60] (0.5+3*\c , 1.7321/4+1.7321*\cc) --(2-0.5*0.75+3*\c, 1.7321/2+1.7321/2+1.7321/8+1.7321*\cc- 1.7321);
  \draw [blue!60] (0.5+3*\c , 1.7321/4+1.7321*\cc) --(2+0.5*0.75+3*\c-3, 1.7321/2+1.7321/2+1.7321/8+1.7321*\cc- 1.7321);
  \draw [blue!60] (2+3*\c , 3*1.7321/4+1.7321*\cc) --(0.5+0.5*0.75+ 3*\c, 1.7321/2+1.7321/8+1.7321*\cc);
 \draw [blue!60] (2+3*\c , 3*1.7321/4+1.7321*\cc) --(0.5-0.5*0.75+ 3*\c+3, 1.7321/2+1.7321/8+1.7321*\cc);
 \draw [blue!60] (2+3*\c , 3*1.7321/4+1.7321*\cc) --(2-0.5*0.75+3*\c, 1.7321/2+1.7321/2+1.7321/8+1.7321*\cc- 1.7321);
  \draw [blue!60] (2+3*\c , 3*1.7321/4+1.7321*\cc) --(2+0.5*0.75+3*\c, 1.7321/2+1.7321/2+1.7321/8+1.7321*\cc- 1.7321); 
   \draw [blue!60] (0.5+0.5*0.75+3*\c, 1.7321/2+1.7321/8+1.7321*\cc) -- (2-0.5*0.75+ 3*\c, 1.7321/2+1.7321/2+1.7321/8+1.7321*\cc- 1.7321);
    \draw [blue!60] (0.5+0.5*0.75+3*\c, 1.7321/2+1.7321/8+1.7321*\cc) -- (2-0.5*0.75+ 3*\c, 1.7321/2+1.7321/2+1.7321/8+1.7321*\cc);
     \draw [blue!60] (2+0.5*0.75+3*\c, 1.7321/2+1.7321/2+1.7321/8+1.7321*\cc) -- (0.5-0.5*0.75+ 3+3*\c, 1.7321/2+1.7321/8+1.7321*\cc);}
     \foreach \cc in {-1,...,3}
     \foreach \c in {-3,...,2}{
     \draw [blue!60] (2+0.5*0.75+3*\c, 1.7321/2+1.7321/2+1.7321/8+1.7321*\cc) -- (0.5-0.5*0.75+ 3+3*\c, 1.7321/2+1.7321/8+1.7321+1.7321*\cc);}
\end{tikzpicture} 
  \caption{Cell partitioning used for the first part of the converse bound \eqref{upphex2}. }
  \label{fig10-1}
 \vspace*{-3ex}
\end{figure}

We start by proving the upper bound 
\begin{equation}
\S^{(F)} + \S^{(S)} \le \frac{M}{2} + \frac{2}{3}\mu_{\Rx} + \frac{4}{3} \mu_{\Tx}.
\end{equation}
 For each power $P>0$, fix a sequence of encoding, cooperation and decoding functions so that the average block power constraint is satisfied and for sufficiently large blocklength $n$, the probability of error does not exceed $\epsilon$. 
Partition the network into red and white cells as depicted in Fig. \ref{fig10-1}. Define
\begin{IEEEeqnarray}{rCl}
\nonumber
\mathbb{M}_{\text{red}} ^{(F)} &:=& \{ M_k^{(F)}: k \text{ is a mobile user in a red cell}\}\\\nonumber
\mathbb{M}_{\text{red}} ^{(S)} &:=& \{ M_k^{(S)}: k \text{ is a mobile user in a red cell}\}\\\nonumber
\mathbb{X}_{\text{red}} &:=& \{ X_k: k \text{ is a mobile user in a red cell}\}\\\nonumber
\mathbb{Y}_{\text{red}} &:= &\{ Y_{\tilde{k}}: {\tilde{k}}\text{ is a red cell}\}\\\nonumber
\mathbb{Y}_{\text{white}} &:= &\{ Y_{\tilde{k}}: {\tilde{k}} \text{ is a white cell}\}\nonumber
\end{IEEEeqnarray}
Fix also an arbitrary pair $(\Dt,\Dr)$ such that $\Dt+\Dr=\D$. Then, for each round $j = 1,\ldots, \Dt$, we introduce the following shortcut for transmitter conferencing messages
\begin{IEEEeqnarray}{rCl}
\nonumber
\mathbb{T}^{(j)}_{C_1 \to C_2}& :=& \{T_{k \to\ell}^{j,(n)} : k \text{ is in a cell colored in}\; C_1 , \\ \nonumber
&&  \hspace{1.4cm}\ell \text{ is  in a cell colored in}\; C_2\}
\end{IEEEeqnarray}
where $k$ and $\ell$ are random mobile users and $C_1$ and $C_2$ are cell colors, i.e., $C_1,C_2\in\{\textnormal{red},  \textnormal{white}\}$. Also, for each $j' = 1,\ldots,\Dr$, define the following shortcuts:
\begin{IEEEeqnarray}{rCl}
\nonumber
\mathbb{Q}^{(j')}_{C_1\to C_2} &:=& \{Q_{\tilde k \to \ell}^{j',(n)} : \tilde k \text{ is a cell colored in}\; C_1, \\ \nonumber 
&& \hspace{1.5cm} \ell \text{ is a cell colored in}\; C_2\}.
\end{IEEEeqnarray}
Consider a virtual super receiver that observes $\mathbb{Y}_{\text{red}}$, $\mathbb{T}_{\text{white} \to \text{red}}$, $\mathbb{Q}_{\text{white} \to \text{red}}$ and genie information $\mathbb{G}$
\begin{equation} \label{yblue}
\mathbb{G} = \mathbb{Y}_{\text{white}}- \mathsf{H}^{-1}_{\text{white} \to \text{red}} (\mathbb{Y}_{\text{red}} - \mathsf{H}_{\text{red} \to \text{red}}\mathbb{X}_{\text{red}}) - \mathsf{H}_{\text{red} \to \text{white}} \mathbb{X}_{\text{red}}, 
\end{equation}
where $\mathsf{H}_{C_1 \to C_2}$ denotes the channel matrix  from the mobile users in the cells of color $C_1$ to the basestations in the cells of color $C_2$. Because these matrices are square and the channel coefficients are drawn i.i.d according to a continues distribution, they are invertible.  Notice that  $\mathbb{G}$  satisfies
\begin{IEEEeqnarray}{rCl} \label{g2}
\lim\limits_{P \to\infty} &&\frac{1}{\log(1+P)} I(M_1,\ldots,M_{3K}; \mathbb{G}| \mathbb{Y}_{\text{red}}, \mathbb{T}^{(1)}_{\text{white} \to \text{red}}, \ldots,\nonumber \\
&& \hspace{1cm}\mathbb{T}^{(\Dt)}_{\text{white} \to \text{red}}, \mathbb{Q}^{(1)}_{\text{white} \to \text{red}},\ldots, \mathbb{Q}^{(\Dr)}_{\text{white} \to \text{red}} )= 0,
\end{IEEEeqnarray}
irrespective of the fixed encoding, cooperation and decoding functions.

~~%
Consider now Algorithm~\ref{alg2}. If the virtual super receiver follows Algorithm \ref{alg2}, then it decodes  all the $3 K$ messages $\{M_k\}$ correctly whenever the $ K$ BSs decode them correctly in the original setup.
\begin{algorithm}[h]
\caption{}
\begin{algorithmic}[1]
\State \textbf {Initialization:}
\For{$j' = 1, \ldots, \Dr$}

\State Apply the cooperation functions $\psi_{\tilde k \to \ell}^{j',(n)}$ to $\mathbb{Y}_{\text{red}}$ and  $\{\mathbb{Q}^{(j'')}_{\text{white} \to \text{red}}\}_{j'' =1}^{j'-1}$ and $\{\mathbb{Q}^{(j'')}_{\text{red} \to \text{red}}\}_{j'' =1}^{j'-1}$.  

\State Compute $\mathbb{Q}^{(j')}_{\text{white} \to \text{red}}$ and $\mathbb{Q}^{(j')}_{\text{red} \to \text{red}}$. 

\EndFor

\State  Apply the decoding function $g_{\tilde k}^{(n)}$ to  $\mathbb{Y}_{\text{red}}$ and $\{\mathbb{Q}^{(j')}_{\text{white} \to \text{red}}\}_{j' =1}^{\Dr}$ and $\{\mathbb{Q}^{(j')}_{\text{red} \to \text{red}}\}_{j' =1}^{\Dr}$ to decode messages $\mathbb M _{\text{red}}$. This yields $\hat {\mathbb M}_{\text{red}}$.

\For{$j = 1, \ldots, \Dt$}

\State Apply the transmitter conferencing functions  $\xi_{k\to \ell}^{j,(n)}$ to  $\hat {\mathbb M}_{\text{red}}$, $\{\mathbb{T}^{(j'')}_{\text{white} \to \text{red}}\}_{j''=1}^{j-1}$ and $\{\mathbb{T}^{(j'')}_{\text{red} \to \text{red}}\}_{j''=1}^{j-1}$. 

\State Compute  $\mathbb{T}^{(j)}_{\text{white} \to \text{red}}$ and $\mathbb{T}^{(j)}_{\text{red} \to \text{red}}$.

\EndFor

\State  Apply the encoding function $f_k^{(n)}$ to the decoded messages $\hat {\mathbb M}_{\text{red}}$, $\{\mathbb{T}^{(j)}_{\text{white} \to \text{red}}\}_{j=1}^{\Dt}$ and $\{\mathbb{T}^{(j)}_{\text{red} \to \text{red}}\}_{j=1}^{\Dt}$ to construct $\mathbb{X}_{\text{red}}$.

\State  Reconstruct $\mathbb{Y}_{\text{white}}$ with  $\mathbb{X}_{\text{red}}$, $\mathbb{Y}_{\text{red}}$, and the genie information  $\mathbb{G}$. 

\For{$j' = 1, \ldots, \Dr$}

\State Apply the  cooperation functions $\psi_{\tilde k \to \ell}^{j',(n)}$ to $\mathbb{Y}_{\text{white}}$, $\mathbb{Y}_{\text{red}}$ and to $\{\mathbb{Q}^{(j'')}_{\text{red} \to \text{white}}\}_{j'' =1}^{j'-1}$ and $\{\mathbb{Q}^{(j'')}_{\text{white} \to \text{white}}\}_{j'' =1}^{j'-1}$. 

\State Compute  $\mathbb{Q}^{(j')}_{\text{red} \to \text{white}}$ and $\mathbb{Q}^{(j')}_{\text{white} \to \text{white}}$.

\EndFor

\State Apply the decoding function $g_{\tilde k}^{(n)}$ to  $\mathbb{Y}_{\text{white}}$ and  $\{\mathbb{Q}^{(j')}_{\text{red} \to \text{white}}\}_{j' =1}^{\Dr}$ and $\{\mathbb{Q}^{(j')}_{\text{white} \to \text{white}}\}_{j'=1}^{\Dr}$ to decode messages $\mathbb M _{\text{white}}$.
\State  \textbf{End}
\end{algorithmic}
\label{alg2}
\end{algorithm}
\par We can therefore conclude that any tuple $(R_1, \ldots, R_{3K})$ with $R_k = R_k^{(F)} + R_k^{(S)}$ and $k = 1, \ldots, 3K$, that is achievable over the original network to the BSs is also achievable over the network to the virtual super receiver. So, by Fano's inequality:
\begin{IEEEeqnarray}{rCl}
		&&3 K(R^{(F)} + R^{(S)}) \le \frac{1}{n} I(M_1,\ldots,M_{3K};\mathbb{Y}_{\text{red}}, \mathbb{T}^{(1)}_{\text{white} \to \text{red}}, \nonumber \\
		&& \ldots,\mathbb{T}^{(\Dt)}_{\text{white} \to \text{red}}, \mathbb{Q}^{(1)}_{\text{white} \to \text{red}}, \ldots,\mathbb{Q}^{(\Dr)}_{\text{white} \to \text{red}},\mathbb{G}) + \frac{\epsilon}{n} 
		\nonumber \\ &=& \frac{1}{n}I(M_1,\ldots,M_{3K};\mathbb{Y}_{\text{red}})
		\nonumber \\ & +& \frac{1}{n}I(M_1,\ldots,M_{3K}; \mathbb{T}^{(1)}_{\text{white} \to \text{red}}, \ldots,\mathbb{T}^{(\Dt)}_{\text{white} \to \text{red}}, \nonumber \\
		&&\hspace{2.5cm}\mathbb{Q}^{(1)}_{\text{white} \to \text{red}}, \ldots,\mathbb{Q}^{(\Dr)}_{\text{white} \to \text{red}}|\mathbb{Y}_{\text{red}})
		\nonumber \\ &+& \frac{1}{n}I(M_1,\ldots,M_{3K}; \mathbb{G}|\mathbb{Y}_{\text{red}},\mathbb{T}^{(1)}_{\text{white} \to \text{red}}, \ldots,\nonumber \\
		&& \mathbb{T}^{(\Dt)}_{\text{white} \to \text{red}}, \mathbb{Q}^{(1)}_{\text{white} \to \text{red}}, \ldots,\mathbb{Q}^{(\Dr)}_{\text{white} \to \text{red}}) + \frac{\epsilon}{n}, \label{secondb}
	\end{IEEEeqnarray}
By considering \eqref{g2} and dividing \eqref{secondb} by $\frac{1}{2}\log(1+P)$ and taking the limits $\epsilon \to 0$ and $P \to \infty$, the following bound is obtained on the multiplexing gain region:
\begin{equation} \label{upH2}
3K (\S^{(F)} + \S^{(S)})  \le 3M|\mathcal{I}_{\text{red}}| + 4( K - |\mathcal{I}_{\text{red}}| )(\mu_{\Rx} + 2 \mu_{\Tx}).
\end{equation}
where $|\mathcal{I}_{\text{red}}|:=\{ i: \text{cell } i \text{ is red} \}$, and 
\begin{equation}\label{Ired}
\lim\limits_{ K \to\infty} \frac{|\mathcal{I}_{\text{red}}|}{ K} = \frac{1}{2}.
\end{equation}
Finally, by considering \eqref{Ired} and dividing \eqref{upH2} by $3K$,  when $ K \to \infty$ the desired bound is established. 
\par Bound \eqref{upphex1} is proved in an analogous way, where since one can restrict to sending only ``fast" messages, one can set  $\mu_{\Tx} = \mu_{\Rx} = 0$ in the proof.
\par We now prove the upper bound 
\begin{equation}
\S^{(F)} + \S^{(S)} \le  M \left (1- \frac{1}{2(1+\D+\D^2)}\right).
\end{equation}
For each power $P>0$, fix a sequence of encoding, cooperation and decoding functions so that the average block power constraint is satisfied and for sufficiently large blocklength $n$, the probability of error does not exceed $\epsilon$. 
{\color{black}
Partition the set of cells $\{1,\ldots, K\}$ into red, pink, blue, and white cells as shown in Fig. \ref{fig5} for the case $\D=3$. 
In particular,  there is an equal number of blue and red cells, and red and blue cells form a regular pattern of equilateral triangles  
that are $\D$ cell-hops apart. Each red cell is surrounded by $6$ pink cells. 
Define
\begin{IEEEeqnarray}{rCl}
|\mathcal{I}_{\text{red}}| &:=&\{ i: \text{cell } i \text{ is red} \} \nonumber\\
|\mathcal{I}_{\text{white}}| &:=&\{ i: \text{cell } i \text{ is white} \} \nonumber\\
|\mathcal{I}_{\text{blue}}| &:=&\{ i: \text{cell } i \text{ is blue} \} \nonumber\\
|\mathcal{I}_{\text{pink}}| &:=&\{ i: \text{cell } i \text{ is pink} \}\nonumber
\end{IEEEeqnarray}
then the following limiting behaviours can be verified:
\begin{IEEEeqnarray}{rCl}
\lim\limits_{ K \to\infty} \frac{|\mathcal{I}_{\text{red}}|}{ K} &=& \frac{1}{2(\D^2 + \D + 1)} \\
\lim\limits_{ K \to\infty} \frac{|\mathcal{I}_{\text{blue}}|}{ K} &=& \frac{1}{2(\D^2 + \D + 1)} \\
\lim\limits_{ K \to\infty} \frac{|\mathcal{I}_{\text{white}}|}{ K} &=& \frac{\D^2 + \D -3}{(\D^2 + \D + 1)} \\
\lim\limits_{ K \to\infty} \frac{|\mathcal{I}_{\text{pink}}|}{ K} &=& \frac{3}{(\D^2 + \D + 1)}. 
\end{IEEEeqnarray}
Consider $\mathbb{T}^{(j)}_{C_1 \to C_2}$ and $\mathbb{Q}^{(j')}_{C_1\to C_2}$ as before and define  
\begin{IEEEeqnarray}{rCl}
\nonumber
\mathbb{M}_{\text{red}} &:=& \{ M_k: k \text{ is a mobile user in a red cell}\}\\\nonumber
\mathbb{X}_{\text{red}} &:=& \{ X_k: k \text{ is a mobile user in a red cell}\}\\\nonumber
\mathbb{Y}_{\text{red}} &:= &\{ Y_i: i \text{ is a red cell}\}\\\nonumber
\mathbb{Y}_{\text{white}} &:= &\{ Y_i: i \text{ is a white cell}\}\\\nonumber
\mathbb{Y}_{\text{blue}} &:= &\{ Y_i: i \text{ is a blue cell}\}\\\nonumber
\mathbb{Y}_{\text{pink}} &:= &\{ Y_i: i \text{ is a pink cell}\}
\end{IEEEeqnarray}
\begin{figure}[t]
  \centering
  \small
 \begin{tikzpicture}[scale=0.35, >=stealth]
 \centering
  \foreach \i in {-2,...,4} 
  \foreach \j in {-1,...,7} {
\draw[ fill = white] (3*\i,2*sin{60}*\j+2*sin{60}) -- +(0:1) -- (3*\i+3*cos{60},2*sin{60}*\j+sin{60}) -- +(120:1) --(3*\i+3*cos{60},2*sin{60}*\j+sin{60}) -- +(-120:1)--(3*\i,2*sin{60}*\j) -- +(0:1)--(3*\i,2*sin{60}*\j) -- +(120:1)-- (3*\i,2*sin{60}*\j+2*sin{60}) -- +(-120:1)  ;}
  \foreach \i in {-2,...,4}
 \foreach \j in {-1,...,7} {
 \draw[ fill= white] (3*\i+3*cos{60},2*sin{60}*\j+sin{60}) -- +(0:1)--(3*\i+3*cos{60},2*sin{60}*\j+sin{60}) -- +(120:1)--(3*\i+3*cos{60},2*sin{60}*\j+3*sin{60}) -- +(-120:1)--(3*\i+3*cos{60},2*sin{60}*\j+3*sin{60}) -- +(0:1)-- (3*\i+3,2*sin{60}*\j+2*sin{60}) -- +(120:1)--(3*\i +3,2*sin{60}*\j+2*sin{60}) -- +(-120:1);  }
   \foreach \i in {0} 
  \foreach \j in {0} {
\draw[ fill = red!60] (3*\i,2*sin{60}*\j+2*sin{60}) -- +(0:1) -- (3*\i+3*cos{60},2*sin{60}*\j+sin{60}) -- +(120:1) --(3*\i+3*cos{60},2*sin{60}*\j+sin{60}) -- +(-120:1)--(3*\i,2*sin{60}*\j) -- +(0:1)--(3*\i,2*sin{60}*\j) -- +(120:1)-- (3*\i,2*sin{60}*\j+2*sin{60}) -- +(-120:1)  ;}
 \foreach \i in {-1} 
  \foreach \j in {7} {
\draw[ fill = red!60] (3*\i,2*sin{60}*\j+2*sin{60}) -- +(0:1) -- (3*\i+3*cos{60},2*sin{60}*\j+sin{60}) -- +(120:1) --(3*\i+3*cos{60},2*sin{60}*\j+sin{60}) -- +(-120:1)--(3*\i,2*sin{60}*\j) -- +(0:1)--(3*\i,2*sin{60}*\j) -- +(120:1)-- (3*\i,2*sin{60}*\j+2*sin{60}) -- +(-120:1)  ;}
  \foreach \i in {-1}
 \foreach \j in {3} {
 \draw[ fill= red!60] (3*\i+3*cos{60},2*sin{60}*\j+sin{60}) -- +(0:1)--(3*\i+3*cos{60},2*sin{60}*\j+sin{60}) -- +(120:1)--(3*\i+3*cos{60},2*sin{60}*\j+3*sin{60}) -- +(-120:1)--(3*\i+3*cos{60},2*sin{60}*\j+3*sin{60}) -- +(0:1)-- (3*\i+3,2*sin{60}*\j+2*sin{60}) -- +(120:1)--(3*\i +3,2*sin{60}*\j+2*sin{60}) -- +(-120:1);}
   \foreach \i in {3} 
  \foreach \j in {5} {
\draw[ fill = red!60] (3*\i,2*sin{60}*\j+2*sin{60}) -- +(0:1) -- (3*\i+3*cos{60},2*sin{60}*\j+sin{60}) -- +(120:1) --(3*\i+3*cos{60},2*sin{60}*\j+sin{60}) -- +(-120:1)--(3*\i,2*sin{60}*\j) -- +(0:1)--(3*\i,2*sin{60}*\j) -- +(120:1)-- (3*\i,2*sin{60}*\j+2*sin{60}) -- +(-120:1)  ;}
  \foreach \i in {3}
 \foreach \j in {1} {
 \draw[ fill= red!60] (3*\i+3*cos{60},2*sin{60}*\j+sin{60}) -- +(0:1)--(3*\i+3*cos{60},2*sin{60}*\j+sin{60}) -- +(120:1)--(3*\i+3*cos{60},2*sin{60}*\j+3*sin{60}) -- +(-120:1)--(3*\i+3*cos{60},2*sin{60}*\j+3*sin{60}) -- +(0:1)-- (3*\i+3,2*sin{60}*\j+2*sin{60}) -- +(120:1)--(3*\i +3,2*sin{60}*\j+2*sin{60}) -- +(-120:1);}
  \foreach \i in {1}
 \foreach \j in {2} {
 \draw[ fill= blue!40] (3*\i+3*cos{60},2*sin{60}*\j+sin{60}) -- +(0:1)--(3*\i+3*cos{60},2*sin{60}*\j+sin{60}) -- +(120:1)--(3*\i+3*cos{60},2*sin{60}*\j+3*sin{60}) -- +(-120:1)--(3*\i+3*cos{60},2*sin{60}*\j+3*sin{60}) -- +(0:1)-- (3*\i+3,2*sin{60}*\j+2*sin{60}) -- +(120:1)--(3*\i +3,2*sin{60}*\j+2*sin{60}) -- +(-120:1);}
    \foreach \i in {1} 
  \foreach \j in {6} {
\draw[ fill = blue!40] (3*\i,2*sin{60}*\j+2*sin{60}) -- +(0:1) -- (3*\i+3*cos{60},2*sin{60}*\j+sin{60}) -- +(120:1) --(3*\i+3*cos{60},2*sin{60}*\j+sin{60}) -- +(-120:1)--(3*\i,2*sin{60}*\j) -- +(0:1)--(3*\i,2*sin{60}*\j) -- +(120:1)-- (3*\i,2*sin{60}*\j+2*sin{60}) -- +(-120:1)  ;}

 \foreach \i in {-2} 
  \foreach \j in {1} {
\draw[ fill = blue!40] (3*\i,2*sin{60}*\j+2*sin{60}) -- +(0:1) -- (3*\i+3*cos{60},2*sin{60}*\j+sin{60}) -- +(120:1) --(3*\i+3*cos{60},2*sin{60}*\j+sin{60}) -- +(-120:1)--(3*\i,2*sin{60}*\j) -- +(0:1)--(3*\i,2*sin{60}*\j) -- +(120:1)-- (3*\i,2*sin{60}*\j+2*sin{60}) -- +(-120:1)  ;}

 \foreach \i in {2} 
  \foreach \j in {-1} {
\draw[ fill = blue!40] (3*\i,2*sin{60}*\j+2*sin{60}) -- +(0:1) -- (3*\i+3*cos{60},2*sin{60}*\j+sin{60}) -- +(120:1) --(3*\i+3*cos{60},2*sin{60}*\j+sin{60}) -- +(-120:1)--(3*\i,2*sin{60}*\j) -- +(0:1)--(3*\i,2*sin{60}*\j) -- +(120:1)-- (3*\i,2*sin{60}*\j+2*sin{60}) -- +(-120:1)  ;}
 \foreach \i in {0} 
  \foreach \j in {-1,1,3,4} {
\draw[ fill = pink] (3*\i,2*sin{60}*\j+2*sin{60}) -- +(0:1) -- (3*\i+3*cos{60},2*sin{60}*\j+sin{60}) -- +(120:1) --(3*\i+3*cos{60},2*sin{60}*\j+sin{60}) -- +(-120:1)--(3*\i,2*sin{60}*\j) -- +(0:1)--(3*\i,2*sin{60}*\j) -- +(120:1)-- (3*\i,2*sin{60}*\j+2*sin{60}) -- +(-120:1)  ;}

\foreach \i in {-1} 
  \foreach \j in {3,4,6} {
\draw[ fill = pink] (3*\i,2*sin{60}*\j+2*sin{60}) -- +(0:1) -- (3*\i+3*cos{60},2*sin{60}*\j+sin{60}) -- +(120:1) --(3*\i+3*cos{60},2*sin{60}*\j+sin{60}) -- +(-120:1)--(3*\i,2*sin{60}*\j) -- +(0:1)--(3*\i,2*sin{60}*\j) -- +(120:1)-- (3*\i,2*sin{60}*\j+2*sin{60}) -- +(-120:1)  ;}

  \foreach \i in {-1}
 \foreach \j in {-1,0,2,4,6,7} {
 \draw[ fill= pink] (3*\i+3*cos{60},2*sin{60}*\j+sin{60}) -- +(0:1)--(3*\i+3*cos{60},2*sin{60}*\j+sin{60}) -- +(120:1)--(3*\i+3*cos{60},2*sin{60}*\j+3*sin{60}) -- +(-120:1)--(3*\i+3*cos{60},2*sin{60}*\j+3*sin{60}) -- +(0:1)-- (3*\i+3,2*sin{60}*\j+2*sin{60}) -- +(120:1)--(3*\i +3,2*sin{60}*\j+2*sin{60}) -- +(-120:1);}
 
  \foreach \i in {-2}
 \foreach \j in {6,7} {
 \draw[ fill= pink] (3*\i+3*cos{60},2*sin{60}*\j+sin{60}) -- +(0:1)--(3*\i+3*cos{60},2*sin{60}*\j+sin{60}) -- +(120:1)--(3*\i+3*cos{60},2*sin{60}*\j+3*sin{60}) -- +(-120:1)--(3*\i+3*cos{60},2*sin{60}*\j+3*sin{60}) -- +(0:1)-- (3*\i+3,2*sin{60}*\j+2*sin{60}) -- +(120:1)--(3*\i +3,2*sin{60}*\j+2*sin{60}) -- +(-120:1);}
 
 \foreach \i in {0}
 \foreach \j in {-1,0} {
 \draw[ fill= pink] (3*\i+3*cos{60},2*sin{60}*\j+sin{60}) -- +(0:1)--(3*\i+3*cos{60},2*sin{60}*\j+sin{60}) -- +(120:1)--(3*\i+3*cos{60},2*sin{60}*\j+3*sin{60}) -- +(-120:1)--(3*\i+3*cos{60},2*sin{60}*\j+3*sin{60}) -- +(0:1)-- (3*\i+3,2*sin{60}*\j+2*sin{60}) -- +(120:1)--(3*\i +3,2*sin{60}*\j+2*sin{60}) -- +(-120:1);}
 
 \foreach \i in {2,3}
 \foreach \j in {4,5} {
 \draw[ fill= pink] (3*\i+3*cos{60},2*sin{60}*\j+sin{60}) -- +(0:1)--(3*\i+3*cos{60},2*sin{60}*\j+sin{60}) -- +(120:1)--(3*\i+3*cos{60},2*sin{60}*\j+3*sin{60}) -- +(-120:1)--(3*\i+3*cos{60},2*sin{60}*\j+3*sin{60}) -- +(0:1)-- (3*\i+3,2*sin{60}*\j+2*sin{60}) -- +(120:1)--(3*\i +3,2*sin{60}*\j+2*sin{60}) -- +(-120:1);}
 
  \foreach \i in {3}
 \foreach \j in {2,0} {
 \draw[ fill= pink] (3*\i+3*cos{60},2*sin{60}*\j+sin{60}) -- +(0:1)--(3*\i+3*cos{60},2*sin{60}*\j+sin{60}) -- +(120:1)--(3*\i+3*cos{60},2*sin{60}*\j+3*sin{60}) -- +(-120:1)--(3*\i+3*cos{60},2*sin{60}*\j+3*sin{60}) -- +(0:1)-- (3*\i+3,2*sin{60}*\j+2*sin{60}) -- +(120:1)--(3*\i +3,2*sin{60}*\j+2*sin{60}) -- +(-120:1);}
 
 \foreach \i in {3} 
  \foreach \j in {2,1,4,6} {
\draw[ fill = pink] (3*\i,2*sin{60}*\j+2*sin{60}) -- +(0:1) -- (3*\i+3*cos{60},2*sin{60}*\j+sin{60}) -- +(120:1) --(3*\i+3*cos{60},2*sin{60}*\j+sin{60}) -- +(-120:1)--(3*\i,2*sin{60}*\j) -- +(0:1)--(3*\i,2*sin{60}*\j) -- +(120:1)-- (3*\i,2*sin{60}*\j+2*sin{60}) -- +(-120:1)  ;}

 \foreach \i in {4} 
  \foreach \j in {2,1} {
\draw[ fill = pink] (3*\i,2*sin{60}*\j+2*sin{60}) -- +(0:1) -- (3*\i+3*cos{60},2*sin{60}*\j+sin{60}) -- +(120:1) --(3*\i+3*cos{60},2*sin{60}*\j+sin{60}) -- +(-120:1)--(3*\i,2*sin{60}*\j) -- +(0:1)--(3*\i,2*sin{60}*\j) -- +(120:1)-- (3*\i,2*sin{60}*\j+2*sin{60}) -- +(-120:1)  ;}
  \foreach \ii in {-1,...,7}
    \foreach \jj in {-2,...,4}{
  \foreach \a in {90,-30,-150} \draw[black,dashed] (0.5+3*\jj,0.8660+\ii*1.7321) -- +(\a:0.8660);
    \foreach \a in {90,-30,-150} \draw[black,dashed] (2+3*\jj,1.7321+\ii*1.7321) -- +(\a:0.8660);}
  \draw[green,thick] (0.5-3*1,0.8660+3*1.7321) -- (0.5-3*2,0.8660+2*1.7321)-- (0.5-3*2,0.8660+4*1.7321)--(0.5-3*1,0.8660+3*1.7321);
   \draw[green,thick] (0.5-3*1,0.8660+4*1.7321) -- (0.5-3*2,0.8660+5*1.7321)-- (0.5-3*1,0.8660+6*1.7321)--(0.5-3*1,0.8660+4*1.7321);
     \draw[green,thick] (0.5-3*0,0.8660+4*1.7321) -- (0.5+3*1,0.8660+5*1.7321)-- (0.5+3*1,0.8660+3*1.7321)--(0.5-3*0,0.8660+4*1.7321);
     \draw[green,thick] (0.5-3*0,0.8660+3*1.7321) -- (0.5+3*1,0.8660+2*1.7321)--(0.5-3*0,0.8660+1*1.7321) --(0.5-3*0,0.8660+3*1.7321);
      \draw[green,thick] (2-3*1,1.7321+2*1.7321)--(2-3*1,1.7321+0*1.7321)--(2-3*2,1.7321+1*1.7321)--(2-3*1,1.7321+2*1.7321);
       \draw[green,thick] (2-3*1,1.7321+4*1.7321)--(2-3*1,1.7321+6*1.7321)--(2-3*0,1.7321+5*1.7321)--(2-3*1,1.7321+4*1.7321);
  \foreach \c in {-2,...,4}
     \foreach \cc in {-1,...,7}{
 \draw [fill=black] (0.5 + 3*\c, 1.7321/4+1.7321*\cc) circle (0.07);}
  \foreach \c in {-2,...,4}
     \foreach \cc in {-1,...,7}{
 \draw [fill=black] (2+ 3*\c, 3*1.7321/4+1.7321*\cc) circle (0.07);}
 \foreach \ccc in {-2,...,4}
     \foreach \cccc in {-1,...,7}{
  \draw [fill=black] (0.5+0.5*0.75+ 3*\ccc, 1.7321/2+1.7321/8+1.7321*\cccc) circle (0.07);
   \draw [fill=black] (2-0.5*0.75+3*\ccc, 1.7321/2+1.7321/2+1.7321/8+1.7321*\cccc) circle (0.07);
     \draw [fill=black] (0.5-0.5*0.75+ 3*\ccc, 1.7321/2+1.7321/8+1.7321*\cccc) circle (0.07);
   \draw [fill=black] (2+0.5*0.75+3*\ccc, 1.7321/2+1.7321/2+1.7321/8+1.7321*\cccc) circle (0.07);}
    \foreach \cc in {0,...,7}
     \foreach \c in {-2,...,4}{
 \draw [blue!60] (0.5+3*\c , 1.7321/4+1.7321*\cc) --(0.5+0.5*0.75+ 3*\c, 1.7321/2+1.7321/8+1.7321*\cc- 1.7321);
 \draw [blue!60] (0.5+3*\c , 1.7321/4+1.7321*\cc) --(0.5-0.5*0.75+ 3*\c, 1.7321/2+1.7321/8+1.7321*\cc- 1.7321);}
 \foreach \cc in {-1,...,7}
     \foreach \c in {-2,...,4}{
 \draw [blue!60] (0.5+3*\c , 1.7321/4+1.7321*\cc) --(2-0.5*0.75+3*\c, 1.7321/2+1.7321/2+1.7321/8+1.7321*\cc- 1.7321);
  \draw [blue!60] (0.5+3*\c , 1.7321/4+1.7321*\cc) --(2+0.5*0.75+3*\c-3, 1.7321/2+1.7321/2+1.7321/8+1.7321*\cc- 1.7321);
  \draw [blue!60] (2+3*\c , 3*1.7321/4+1.7321*\cc) --(0.5+0.5*0.75+ 3*\c, 1.7321/2+1.7321/8+1.7321*\cc);
 \draw [blue!60] (2+3*\c , 3*1.7321/4+1.7321*\cc) --(0.5-0.5*0.75+ 3*\c+3, 1.7321/2+1.7321/8+1.7321*\cc);
 \draw [blue!60] (2+3*\c , 3*1.7321/4+1.7321*\cc) --(2-0.5*0.75+3*\c, 1.7321/2+1.7321/2+1.7321/8+1.7321*\cc- 1.7321);
  \draw [blue!60] (2+3*\c , 3*1.7321/4+1.7321*\cc) --(2+0.5*0.75+3*\c, 1.7321/2+1.7321/2+1.7321/8+1.7321*\cc- 1.7321); 
   \draw [blue!60] (0.5+0.5*0.75+3*\c, 1.7321/2+1.7321/8+1.7321*\cc) -- (2-0.5*0.75+ 3*\c, 1.7321/2+1.7321/2+1.7321/8+1.7321*\cc- 1.7321);
    \draw [blue!60] (0.5+0.5*0.75+3*\c, 1.7321/2+1.7321/8+1.7321*\cc) -- (2-0.5*0.75+ 3*\c, 1.7321/2+1.7321/2+1.7321/8+1.7321*\cc);
     \draw [blue!60] (2+0.5*0.75+3*\c, 1.7321/2+1.7321/2+1.7321/8+1.7321*\cc) -- (0.5-0.5*0.75+ 3+3*\c, 1.7321/2+1.7321/8+1.7321*\cc);}
     \foreach \cc in {-2,...,6}
     \foreach \c in {-3,...,4}{
     \draw [blue!60] (2+0.5*0.75+3*\c, 1.7321/2+1.7321/2+1.7321/8+1.7321*\cc) -- (0.5-0.5*0.75+ 3+3*\c, 1.7321/2+1.7321/8+1.7321+1.7321*\cc);}
   
\end{tikzpicture} 
  \caption{Cell partitioning for $\D = 3$, used for the second part of the converse bound \eqref{upphex2}.  }
  \label{fig5}
 \vspace*{-3ex}
\end{figure}
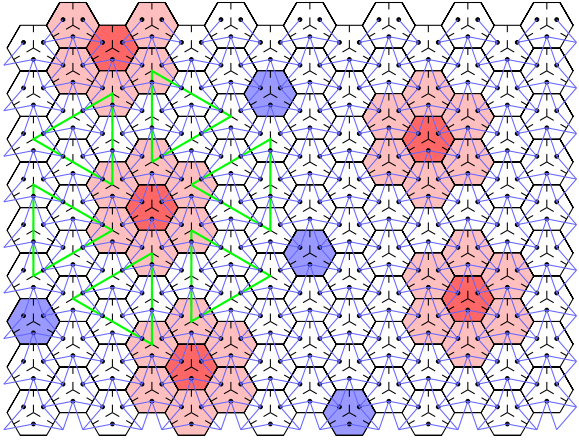~~%
\begin{algorithm}[t]
\caption{}
\begin{algorithmic}[1]
\State \textbf {Initialization:}
\For {$j' = 1, \ldots, \Dr$}
\State Apply the cooperation functions $\psi_{\tilde k \to \ell}^{j',(n)}$ to $\mathbb{Y}_{\text{red}}$,$\mathbb{Y}_{\text{pink}}$, $\mathbb{Y}_{\text{white}}$, and $\{\mathbb{Q}^{(j'')}_{\text{white} \to \text{pink}}\}_{j'' =1}^{j'-1}$, $\{\mathbb{Q}^{(j'')}_{\text{pink} \to \text{white}}\}_{j'' =1}^{j'-1}$, $\{\mathbb{Q}^{(j'')}_{\text{red} \to \text{pink}}\}_{j'' =1}^{j'-1}$, $\{\mathbb{Q}^{(j'')}_{\text{pink} \to \text{red}}\}_{j'' =1}^{j'-1}$, $\{\mathbb{Q}^{(j'')}_{\text{pink} \to \text{pink}}\}_{j'' =1}^{j'-1}$ and $\{\mathbb{\bar Q}^{(j'')}_{\text{white} \to \text{white}}\}_{j'' =1}^{j'-1}$.

\State Compute  $\mathbb{Q}^{(j')}_{\text{white} \to \text{pink}}$, $\mathbb{Q}^{(j')}_{\text{pink} \to \text{white}}$, $\mathbb{Q}^{(j')}_{\text{red} \to \text{pink}}$, $\mathbb{Q}^{(j')}_{\text{pink} \to \text{red}}$, $\mathbb{Q}^{(j')}_{\text{pink} \to \text{pink}}$ and $\mathbb{Q}^{(j')}_{\text{white} \to \text{white}}$.
\EndFor

\State Apply the  decoding functions $g_{\tilde k}^{(n)}$ to  $\mathbb{Y}_{\text{red}}$ and $\{\mathbb{Q}^{(j')}_{\text{pink} \to \text{red}}\}_{j' =1}^{\Dr}$  to decode messages $\mathbb M _{\text{red}}$. This yields $\hat {\mathbb M}_{\text{red}}$.

\For {$j = 1, \ldots, \Dt$}
\State Apply the  conferencing functions  $\xi_{k\to \ell}^{j,(n)}$ to $\hat {\mathbb M}_{\text{red}}$ and  $\{\mathbb{T}^{(j'')}_{\text{white} \to \text{pink}}\}_{j''=1}^{j-1}$,  $\{\mathbb{T}^{(j'')}_{\text{pink} \to \text{white}}\}_{j''=1}^{j-1}$,  $\{\mathbb{T}^{(j'')}_{\text{red} \to \text{pink}}\}_{j''=1}^{j-1}$,  $\{\mathbb{T}^{(j'')}_{\text{pink} \to \text{red}}\}_{j''=1}^{j-1}$,  $\{\mathbb{T}^{(j'')}_{\text{pink} \to \text{pink}}\}_{j''=1}^{j-1}$ and $\{\mathbb{ T}^{(j'')}_{\text{white} \to \text{white}}\}_{j''=1}^{j-1}$.

\State Compute $\mathbb{T}^{(j)}_{\text{white} \to \text{pink}}$, $\mathbb{T}^{(j)}_{\text{pink} \to \text{white}}$, $\mathbb{T}^{(j)}_{\text{red} \to \text{pink}}$, $\mathbb{T}^{(j)}_{\text{pink} \to \text{red}}$, $\mathbb{T}^{(j)}_{\text{pink} \to \text{pink}}$ and $\mathbb{T}^{(j)}_{\text{white} \to \text{white}}$. 
\EndFor

\State Apply the encoding function $f_k^{(n)}$ to  $\hat {\mathbb M}_{\text{red}}$, $\{\mathbb{T}^{(j)}_{\text{pink} \to \text{red}}\}_{j=1}^{\Dt}$  to construct $\mathbb{X}_{\text{red}}$. 

\State Reconstruct $\mathbb{Y}_{\text{blue}}$ with $\mathbb{X}_{\text{red}}$, $\mathbb{Y}_{\text{red}}$, $\mathbb{Y}_{\text{pink}}$ and $\mathbb{Y}_{\text{white}}$, and the genie information  $\mathbb{G}$. 

\For {$j = 1,\ldots, \Dr$}
\State Apply the  cooperation functions $\psi_{\tilde k \to \ell}^{j',(n)}$ to  $\mathbb{Y}_{\text{red}}$, $\mathbb{Y}_{\text{pink}}$, $\mathbb{Y}_{\text{white}}$ and $\mathbb{Y}_{\text{blue}}$, and to the previously calculated receiver conferencing messages.

\State Compute \emph{all} round-$j'$ receiver conferencing messages. 

\EndFor

\State Apply the appropriate decoding function $g_{\tilde k}^{(n)}$ to the output signals  $\mathbb{Y}_{\text{pink}}$, $\mathbb{Y}_{\text{white}}$ and $\mathbb{Y}_{\text{blue}}$, and the required conferencing messages so as to decode messages $\mathbb M _{\text{pink}}$, $\mathbb M _{\text{white}}$ and $\mathbb M _{\text{blue}}$.

\State  \textbf{End}
\end{algorithmic}
\label{alg3}
\end{algorithm}
\par Consider a virtual  super receiver that observes $\mathbb{Y}_{\text{red}}$, $\mathbb{Y}_{\text{pink}}$, $\mathbb{Y}_{\text{white}}$ and the genie information $\mathbb{G}$ defined in \eqref{yblue}, which satisfies
\begin{equation} \label{g3}
\lim\limits_{P \to\infty} \frac{I(M_1,\ldots,M_{3K}; \mathbb{G}| \mathbb{Y}_{\text{red}},\mathbb{Y}_{\text{pink}},\mathbb{Y}_{\text{white}} )}{\log(1+P)} = 0
\end{equation}
irrespective of the fixed encoding, cooperation and decoding functions.
%
%
%
If the virtual super receiver follows Algorithm \ref{alg3}, then it decodes all the $3K$ messages $\{M_k\}$ correctly whenever the $ K$ BSs decode them correctly in the original setup. So, by  Fano's inequality:
\begin{IEEEeqnarray}{rCl}
\lefteqn{	3	K(R^{(F)}+ R^{(S)})} \quad \nonumber \\
&\le &\frac{1}{n} I(M_1,\ldots,M_{3K};\mathbb{Y}_{\text{red}}, \mathbb{Y}_{\text{pink}},\mathbb{Y}_{\text{white}}, \mathbb{G})  + \frac{\epsilon}{n}\nonumber \\
&=&\frac{1}{n}I(M_1,\ldots,M_{3K};\mathbb{Y}_{\text{red}}, \mathbb{Y}_{\text{pink}},\mathbb{Y}_{\text{white}}) \nonumber \\
		& &+ \frac{1}{n}I(M_1,\ldots,M_{3K}; \mathbb{G}|\mathbb{Y}_{\text{red}}, \mathbb{Y}_{\text{pink}},\mathbb{Y}_{\text{white}}) + \frac{\epsilon}{n}, \label{thirdb}
	\end{IEEEeqnarray}
	where $\epsilon$ tends to 0 as $n\to \infty$ and the probability of error in the original setup tends to 0.
Considering \eqref{g3}, dividing \eqref{thirdb}  by $\frac{1}{2}\log(1+P)$, and taking the limits $\epsilon \to 0$, $n\to \infty$, and $P \to \infty$, the following bound is obtained on the multiplexing gain region:
\begin{equation}
3K (\S^{(F)} + \S^{(S)}) \le 3M(|\mathcal{I}_{\text{red}}| + |\mathcal{I}_{\text{pink}}| + |\mathcal{I}_{\text{white}}|)
\end{equation}
Dividing by $3K$ and letting $ K \to \infty$ the desired bound is established.}
\section{Summary and Conclusion}
We bounded the multiplexing gain region of the sectorized hexagonal network when both mobile users and base stations can cooperation and messages are subject to mixed delay constraints. Our results  show that for small cooperation rates, the sum-multiplexing gain is not decreased by the  stringent delay constraint on the ``fast" messages. Networks with mixed delay constraints are particularly timely in view of IoT applications such as automatic traffic control and real-time monitoring. New coding schemes that can efficiently use resources (space/time/frequency) under this framework are thus of high practical relevance.

\section*{Acknowledgement}
The work of H. Nikbakth and M. Wigger was supported by the ERC under grant agreement no. 715111. The work of S. Shamai has been supported by the European Union's Horizon 2020 Research And Innovation Programme, grant agreement no. 694630.

\end{document}